\documentclass[12pt,a4paper]{article}

\usepackage{epsf}
\usepackage{latexsym,amssymb,euscript}
\usepackage[dvips]{graphicx}
\usepackage[numbers,sort&compress]{natbib}
\usepackage{amsmath}
\usepackage{slashed}
\usepackage{booktabs}
\usepackage{hyperref}
\usepackage{braket}
\usepackage{bbold}
\usepackage{graphics}
\usepackage{graphicx}
\usepackage{caption}
\usepackage{subcaption}
\usepackage[titletoc]{appendix}
\graphicspath{{./figures/}}
\hypersetup{
 linktocpage = true,
 linkcolor = blue,
 urlcolor = blue,
 colorlinks = true,
 linkcolor = blue,
 anchorcolor = blue,
 citecolor = blue,
 urlcolor = blue,
 pdfstartview = {XYZ null null 1.25} 
           }
\usepackage[left=3cm, right=2cm]{geometry}
\usepackage{pstricks}
\usepackage{color}
\usepackage{float}

\newcommand{\tarr}{
\begin{array}}
\newcommand{\earr}{\end{array}}

\begin{document}

\begin{titlepage}

\begin{center}
  {\LARGE \bf High-energy resummed distributions \\ for the inclusive
    Higgs-plus-jet production \\ at the LHC}
\end{center}

\vskip 0.5cm

\begin{center}
Francesco~G.~Celiberto~$^{1,2,3,4*}$,
Dmitry Yu.~Ivanov~$^{5,6\S}$,\\
Mohammed~M.A.~Mohammed~$^{7,8\dagger}$,
and Alessandro~Papa~$^{7,8\ddagger}$
\end{center}

\vskip .6cm

\centerline{${}^1$ {\sl Dipartimento di Fisica, Universit\`a degli Studi di
    Pavia, I-27100 Pavia, Italy}}
\vskip .2cm
\centerline{${}^2$ {\sl Istituto Nazionale di Fisica Nucleare, Sezione di Pavia,
    I-27100 Pavia, Italy}}
\vskip .2cm
\centerline{${}^3$ {\sl European Centre for Theoretical Studies
      in Nuclear Physics and Related Areas (ECT*),}}
\centerline{{\sl I-38123 Villazzano, Trento, Italy}}
\vskip .2cm
\centerline{${}^4$ {\sl Fondazione Bruno Kessler (FBK), I-38123
      Povo, Trento, Italy}}
\vskip .2cm
\centerline{${}^5$ {\sl Sobolev Institute of Mathematics, 630090 Novosibirsk,
   Russia}}
\vskip .2cm
\centerline{${}^6$ {\sl Novosibirsk State University, 630090 Novosibirsk,
    Russia}}
\vskip .2cm
\centerline{${}^7$ {\sl Dipartimento di Fisica, Universit\`a della Calabria}}
\centerline{\sl I-87036 Arcavacata di Rende, Cosenza, Italy}
\vskip .2cm
\centerline{${}^8$ {\sl Istituto Nazionale di Fisica Nucleare, Gruppo collegato
      di Cosenza}}
\centerline{\sl I-87036 Arcavacata di Rende, Cosenza, Italy}
\vskip 2cm

\begin{abstract}
  The inclusive hadroproduction of a Higgs boson and of a jet, featuring large
  transverse momenta and well separated in rapidity, is proposed as a novel
  probe channel for the manifestation of the Balitsky-Fadin-Kuraev-Lipatov
  (BFKL) dynamics. Using the standard BFKL approach, with partial inclusion
  of next-to-leading order effects, predictions are presented for azimuthal
  Higgs-jet correlations and other observables, to be possibly compared
  with experimental analyses at the LHC and with theoretical predictions
  obtained in different schemes.
\end{abstract}

\vskip .5cm

\vfill

$^{*}${\it e-mail}:
\href{mailto:francescogiovanni.celiberto@unipv.it}{francescogiovanni.celiberto@unipv.it}

$^{\S}${\it e-mail}:
\href{mailto:d-ivanov@math.nsc.ru}{d-ivanov@math.nsc.ru}

$^{\dagger}${\it e-mail}:
\href{mailto:mohammed.maher@unical.it}{mohammed.maher@unical.it}

$^{\ddagger}${\it e-mail}:
\href{mailto:alessandro.papa@fis.unical.it}{alessandro.papa@fis.unical.it}

\end{titlepage}


\section{Introduction}
\label{introduction}

The Balitsky-Fadin-Kuraev-Lipatov (BFKL)~\cite{BFKL} approach is a consistent
framework for the theoretical study in perturbative QCD of \emph{semi-hard}
processes~\cite{Gribov:1984tu}, where the scale hierarchy, $s \gg Q^2 \gg
\Lambda_{\rm QCD}^2$ holds, $s$ being the squared center-of-mass energy, $Q$
the hard scale given by the process kinematics and $\Lambda_{\rm QCD}$ the QCD
mass scale. For these processes, large energy logarithms enter the
perturbative series with a power increasing with the perturbative order and
compensate thereby the smallness of the strong coupling, $\alpha_s$, thus
calling for an all-order resummation. Within the BFKL approach, such a 
resummation is now amenable both in leading (LLA) and next-to-leading
(NLA) approximations, and some steps towards the extension of the formalism
beyond the NLA have also been done (see, {\it e.g.} Ref.~\cite{Fadin:2016wso}).

In the BFKL framework, the cross section of hadronic processes takes a
peculiar factorized form, combining two impact factors, related to the
transition from each colliding particle to the final-state object produced in
the respective fragmentation region, and a process-independent Green's function.
The latter is determined by an integral equation, whose kernel is known at the
next-to-leading order (NLO) both for forward scattering ({\it i.e.} for $t=0$
and color singlet in the in the
$t$-channel)~\cite{Fadin:1998py,Ciafaloni:1998gs} and for any fixed, not
growing with $s$, momentum transfer $t$ and any possible two-gluon color
state in the $t$-channel~\cite{Fadin:1998jv,FG00,FF05}.

Unfortunately, the list of impact factors known in the NLO is very short:
1) colliding-parton (quarks and gluons) impact
factors~\cite{fading,fadinq,Cia,Ciafaloni:2000sq}, which represent the common
basis for the calculation of the 2) forward-jet impact
factor~\cite{bar1,bar2,Caporale:2011cc,Ivanov:2012ms,Colferai:2015zfa} and of
the 3) forward light-charged hadron one~\cite{Ivanov:2012iv}, 4) the impact
factor describing the $\gamma^*$ to light-vector-meson leading twist
transition~\cite{IKP04}, and 5) the $\gamma^*$ to $\gamma^*$
transition~\cite{gammaIF,Balitsky2012}. This limits considerably the number
of reactions which can be studied fully in the NLA BFKL approach. To enlarge
this number, one has to resort to partial inclusion of NLA effects, by taking
the two impact factors, or just one of them, in the leading-order (LO), using
though the NLA BFKL Green's function.

Putting together full and partial NLA analyses, a respectable number of
semi-hard reactions have been studied so far (see Refs.~\cite{Celiberto:2017ius,Celiberto:2020wpk}
for a review): the diffractive leptoproduction of two light vector
mesons~\cite{Ivanov:2004pp,Ivanov:2005gn,Ivanov:2006gt,Enberg:2005eq},
the total cross section of two highly-virtual photons~\cite{Ivanov:2014hpa},
the inclusive hadroproduction of two jets featuring large transverse momenta
and well separated in rapidity (Mueller-Navelet channel~\cite{Mueller:1986ey}),
for which several phenomenological studies have appeared so
far~\cite{Marquet:2007xx,Colferai:2010wu,Caporale:2012ih,Ducloue:2013wmi,Ducloue:2013bva,Caporale:2013uva,Ducloue:2014koa,Caporale:2014gpa,Ducloue:2015jba,Caporale:2015uva,Celiberto:2015yba,Celiberto:2015mpa,Celiberto:2016ygs,Celiberto:2016vva,Caporale:2018qnm,Chachamis:2015crx,Colferai:2016inu},
the inclusive detection of two light-charged hadrons~\cite{Celiberto:2016hae,Celiberto:2016zgb,Celiberto:2017ptm}, three- and four-jet
hadroproduction~\cite{Caporale:2015vya,Caporale:2015int,Caporale:2016soq,Caporale:2016vxt,Caporale:2016xku,Celiberto:2016vhn,Caporale:2016djm,Caporale:2016lnh,Caporale:2016zkc}, $J/\Psi$-jet~\cite{Boussarie:2017oae},
hadron-jet~\cite{Bolognino:2018oth,Bolognino:2019yqj,Bolognino:2019cac,Celiberto:2020rxb},
Drell-Yan--jet~\cite{Golec-Biernat:2018kem,Deak:2018obv} and heavy-quark pair
photo-~\cite{Celiberto:2017nyx,Bolognino:2019ouc} and
hadroproduction~\cite{Bolognino:2019yls}.

Another engaging direction is represented by the possibility of probing the
proton structure at low-$x$ through the BFKL resummation.
More in particular, the emission of a single forward particle in lepton-proton
or proton-proton scatterings offers us the chance to define an
\emph{unintegrated gluon distribution} (UGD) in the proton, written as a
suitable convolution of the BFKL gluon Green's function and of a
non-perturbative proton impact factor.
Formerly used for the investigation of DIS structure
functions~\cite{Hentschinski:2012kr}, the UGD has later been studied via the
exclusive diffractive electroproduction of a single light vector meson~\cite{Besse:2013muy,Bolognino:2018rhb,Bolognino:2018mlw,Bolognino:2019bko,Celiberto:2019slj,Bolognino:2019pba} at HERA and via the forward inclusive
Drell-Yan production~\cite{Motyka:2014lya,Brzeminski:2016lwh,Celiberto:2018muu}
at LHCb.
Then, determinations of \emph{collinear parton distribution functions} (PDFs) at
NLO and next-to-NLO (NNLO) fixed-order calculations, improved via the inclusion of
NLA small-$x$ effects, were proposed in the last
years~\cite{Ball:2017otu,Abdolmaleki:2018jln,Bonvini:2019wxf}. Quite recently,
a model calculation of unpolarized and polarized
\emph{transverse-momentum-dependent} (TMD) gluon distributions effectively
encoding a BFKL-driven input on small-$x$ tails was
performed~\cite{Bacchetta:2020vty}.

\begin{figure}[t]
\centering
\includegraphics[width=0.55\textwidth]{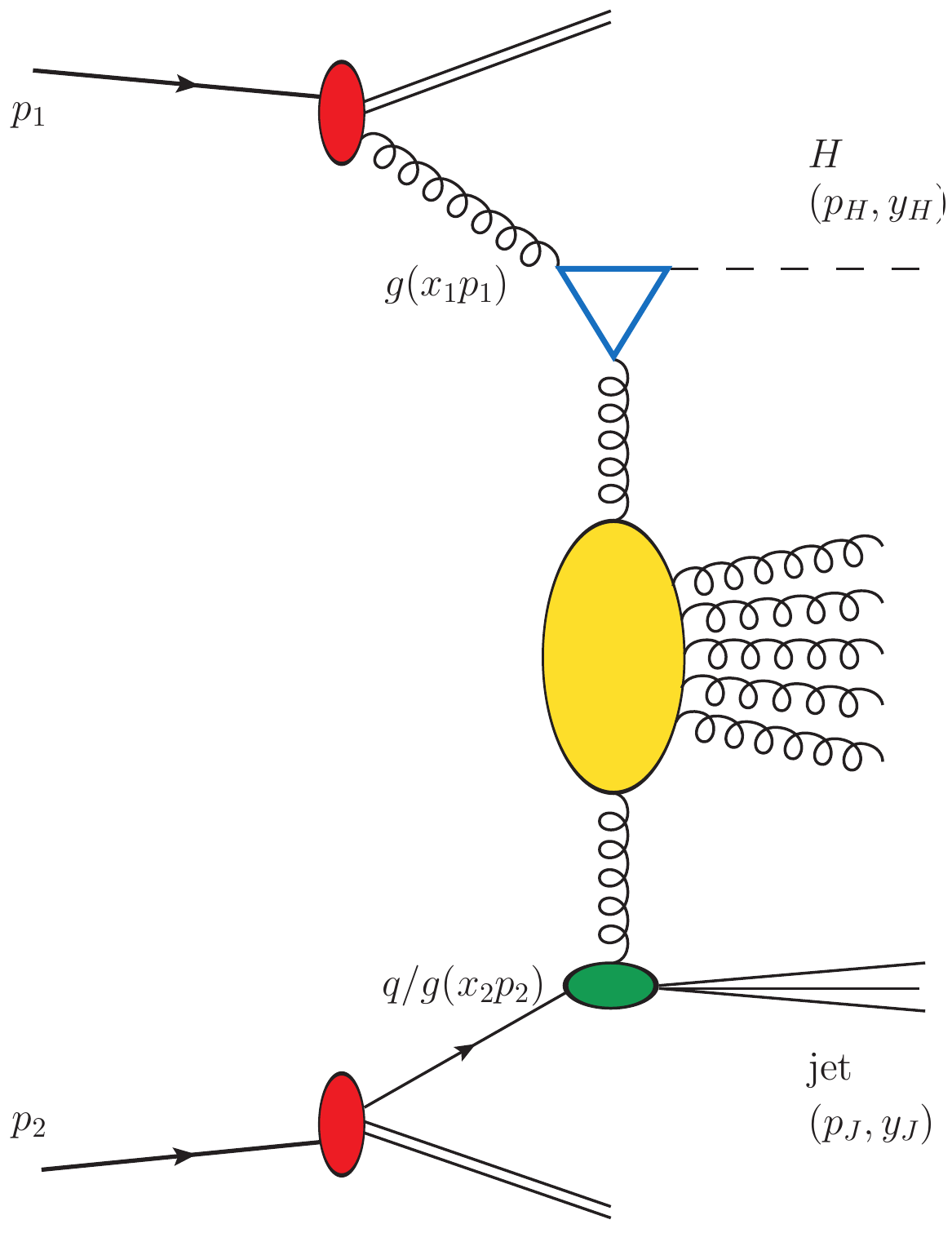}
\caption{Schematic representation of the inclusive Higgs-jet hadroproduction.}
\label{fig:process}
\end{figure}

In this work we introduce and study with NLA BFKL accuracy a novel semi-hard
reaction, \emph{i.e.} the concurrent inclusive production of a Higgs boson
and a jet:
\begin{equation}
\label{process}
{\rm proton}(p_1) \ + \ {\rm proton}(p_2) \ \to \ H(\vec p_H, y_H) \ + \ {\rm X} \ + \ {\rm jet}(\vec p_J, y_J) \;,
\end{equation}
emitted with large transverse momenta, $\vec p_{H,J} \gg \Lambda_{\rm QCD}$, and
separated by a large rapidity gap, $\Delta Y = y_H - y_J$. In
Fig.~\ref{fig:process} we present a pictorial view of this process, in the case
when the tagged object in the forward (backward) rapidity region is the Higgs
boson (jet).

For a Higgs boson with mass $M_H = 125.18$ GeV, the longitudinal-momentum
fraction of the parent proton carried by the struck gluon is rather small,
down to $x \sim 10^{-4} \div 10^{-3}$, making it possible to give a description
at the hand of the BFKL resummation. Recently, a systematic framework to
implement both the BFKL and the Sudakov resummations for the Higgs boson plus
jet production~\cite{Xiao:2018esv}, based on the TMD factorization, has been
developed. The dominant partonic subprocess for
the inclusive Higgs production at the LHC is represented by the gluon-gluon
fusion, $gg \to H$, where the Higgs couples to gluons via a (top) quark loop,
with coupling proportional to the (top) quark mass $M_t$.
In our proposal, following Ref.~\cite{DelDuca:1993ga}, we adopt a kinematics
which strictly respects the semi-hard regime, with the hard-scale set by the
Higgs and top-quark masses, and the Higgs and jet transverse momenta
satisfying the condition, $p_{H,J}^2 \simeq M_H^2$. Moreover, to avoid the
appearance of Sudakov double logarithms (see, {\it e.g.},
Ref.~\cite{Mueller:2015ael}), unraveling only the high-energy ones,
we introduce suitable cuts on transverse momenta to prevent the back-to-back
emission of the Higgs and the jet, or to make this kinematical region marginal
with respect to the remaining phase space. The tag of a jet in the peripheral
regions of the detectors insures the existence of a large rapidity interval,
$\Delta Y = y_H - y_J \simeq \ln(s/Q^2)$, with $Q^2$ a typical hard-scale value.
In our analysis we do not include the resummation of Higgs transverse-momentum logarithms, which has been considered in several recent studies~\cite{Monni:2016ktx,Bizon:2017rah,Bizon:2018foh,Chen:2018pzu,Monni:2019yyr}.

The key ingredient, needed for the study of our process in the BFKL approach,
is the impact factor portraying the transition from a parton to a
final-state Higgs boson, in the scattering off a Reggeized gluon. At the LO,
the initial-state parton can only be a gluon~\footnote{Recently,
    the Higgs impact factor has been calculated also to the
    NLO~\cite{Hentschinski:2020tbi}.}.

We will give predictions for cross section and correlations between the
azimuthal angles of the Higgs and the jet in a theoretical setup where
NLA BFKL effects are included at the level of the Green's function.

The motivation for this work is twofold: on the phenomenological side, we want
to calculate the cross section and to study the angular distributions of the
process~\ref{fig:process} at LHC energies. Note that for that case
the final-state objects to be identified are within current experimental
reach of the LHC; in particular, the detection of the Higgs can profit by the
well tried tools developed for its discovery. On the theoretical side, our
approach is in a sense complementary to the most common ones devoted to
Higgs production, where high-energy (or small-$x$) effects are possibly
included as an improvement with respect to fixed-order calculations in
collinear-factorization (see, {\it e.g.},
Refs.~\cite{Bonvini:2018ixe,Bonvini:2018iwt} where the
Altarelli-Ball-Forte (ABF) small-$x$ resummation formalism is
adopted~\cite{ABF}, and Ref.~\cite{Hautmann:2002tu}). Here, the view is reversed: we consider just
high-energy effects, in the kinematical range where they only matter.
Our results, which are the first for this kind of process encoding NLA
BFKL effects, can therefore be used as a term of comparison for the other
approaches, and contribute thereby to an improvement of our understanding
of strong interactions. Moreover, the notorious problem of the NLA BFKL
corrections, {\it i.e.} that they are large and opposite in sign with respect
to the LLA, should not affect severely the determination of azimuthal
correlations\footnote{For ratios of azimuthal correlations it was shown that
NLO effects are generally milder~\cite{Vera:2006un,Vera:2007kn}.}, due to
the large energy scale provided by the Higgs mass.

The main theoretical limitation of the present work is that the impact factor
for the Higgs production is taken at the LO, although, as explained later,
some NLO terms predictable on the basis of renormalization group analysis,
have been included in our calculation. This may seem reductive, especially
in consideration that Higgs-plus-jet production was already calculated
in QCD at the NLO~\cite{Bonciani:2016qxi,Jones:2018hbb} and even in 
NNLO
QCD~\cite{Boughezal:2013uia,Chen:2014gva,Boughezal:2015aha,Boughezal:2015dra}
through the Higgs effective field theory (HEFT)~\cite{Wilczek:1977zn}.
We believe that this limitation does not spoil the global picture, since in
the high-energy limit the NLA effects in the BFKL Green's function dominate
over those in the impact factors. Nonetheless, the inclusion of NLO corrections
to the Higgs impact factor is doable, though not trivial, and could be
considered in future publications.

The paper is organized as follows: 
Section~\ref{theory} is to set the theoretical framework up;
Section~\ref{phenomenology} is devoted to our results for cross sections and
azimuthal-angle correlations as a function of the rapidity interval,
$\Delta Y$, between the tagged final objects (the Higgs boson and the jet);
Section~\ref{conclusions} carries our closing statements and some outlook.

\section{Theoretical framework}
\label{theory}

\begin{figure}[t]
\centering
\includegraphics[width=0.40\textwidth]{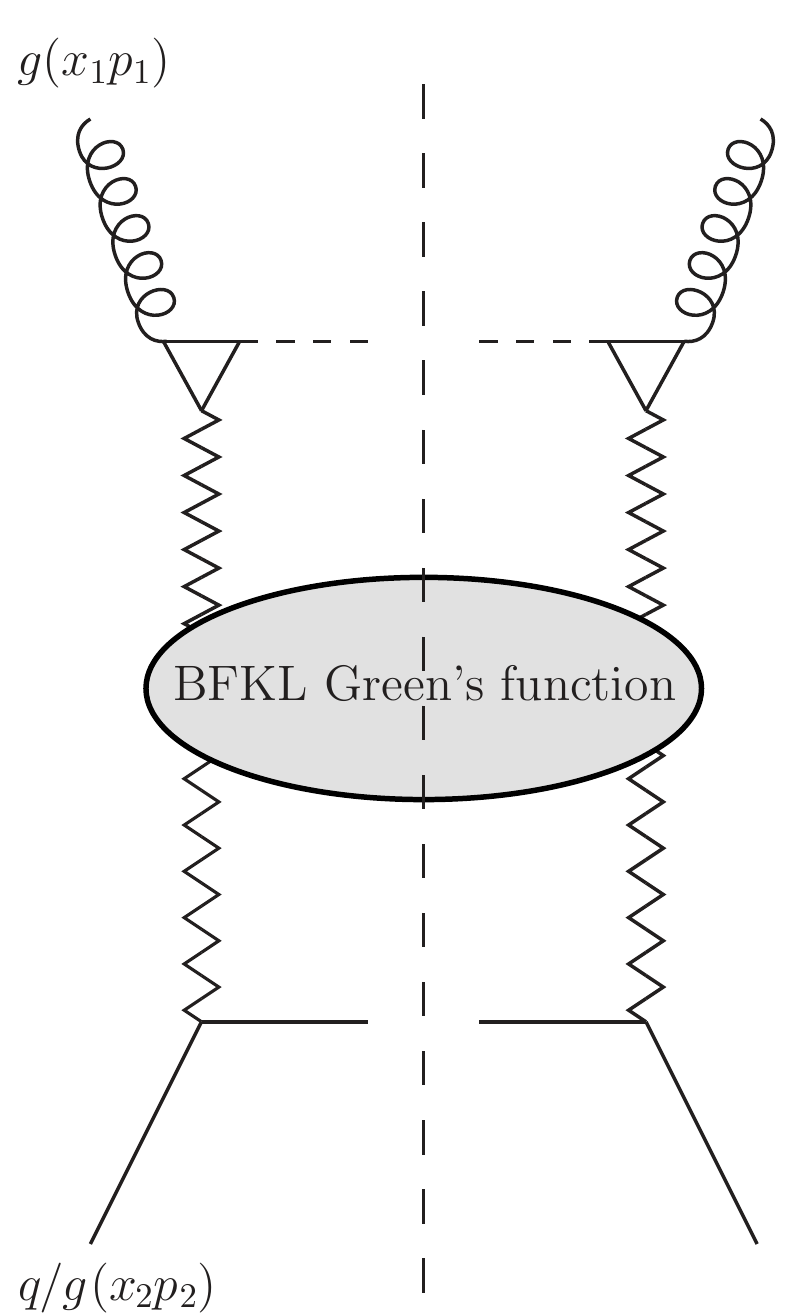}
\caption{Schematic representation of the BFKL factorization for the Higgs-jet
  hadroproduction.}
\label{fig:BFKL_factorization}
\end{figure}

For the process under consideration (see Fig.~\ref{fig:process}) we plan to
construct the cross section, differential in some of the kinematic variables of
the Higgs and the jet, and some azimuthal correlations between them. In the
BFKL approach the cross section takes the factorized form, diagrammatically
represented in Fig.~\ref{fig:BFKL_factorization}, given by the convolution of
the Higgs and jet impact factors with the BFKL gluon Green's function, $G$.

\subsection{Forward-Higgs LO impact factor}
\label{impact_factor}

\begin{figure}[t]
\centering
\includegraphics[width=0.75\textwidth]{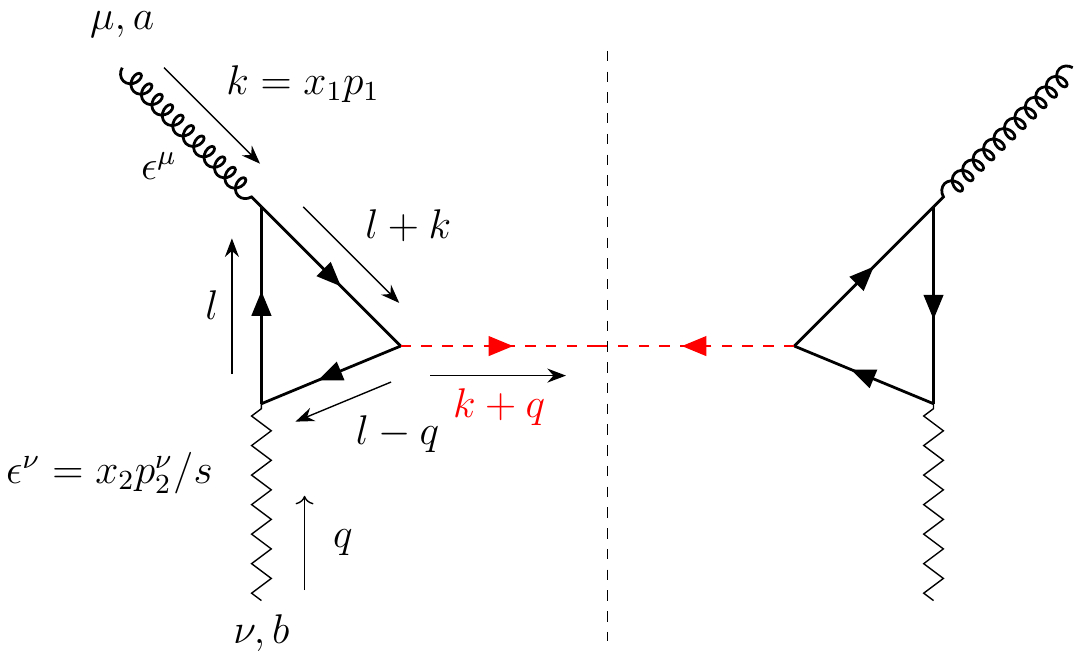}
\caption{Representative Feynman diagram for the squared modulus of the
  amplitude for the gluon scattering off a Reggeon to produce a Higgs particle.
  The Reggeized gluon is depicted by the zigzag line.}
\label{fig:optical_theorem}
\end{figure}

We can define the LO impact factor for the production of the Higgs in the
gluon-gluon fusion channel, as follows (see, {\it e.g.},
Ref.~\cite{Fadin:1998fv}):
\begin{equation}\label{IF_def_M}
 V^{(0)}_{g\to H}(\vec {q})=\sum_{\left\{f\right\}}\int  
 \frac{ds_{gR}}{2\pi} d({\rm PS})^{(f)} |\mathcal{M}|^2 \, ,
\end{equation}
where $\mathcal{M}$ is the amplitude for the scattering of a gluon $g$, emitted
by the colliding proton, off a Reggeon $R$ to produce a final state $f$, which
at the LO, can only consist in a Higgs particle (see
Fig.~\ref{fig:optical_theorem} for a representation of $|\mathcal{M}|^2$).
The integration over the phase space $d({\rm PS})^{(f)}$ then simply
gives
\begin{equation}\label{IF_PS}
  {\rm PS}^{(1)}=\int\frac{d^{4}p_{H}}{(2\pi)^{4}} (2 \pi)
  \delta(p^2_{H}-M_H^2)(2\pi)^{4}
  \delta^{(4)}(k+q-p_{H})= (2\pi) \delta(s_{gR}-M_H^2) \; ,
\end{equation} 
where $p_{H}$ is the Higgs boson momentum. Using this result, we end up with  
\begin{equation}\label{IF_partonic}
  V^{(0)}_{g\to H}(\vec {q})= \frac{\alpha_s^2}{v^2}\frac{\vec q^{\: 2}|
    \mathcal{F}(\vec q^{\: 2})|^2}  
 {128\pi^2\sqrt{N^2_{c}-1}} \; ,
\end{equation}
with $V^{(0)}_{g\to H}(\vec {q})|_{\vec {q}=0}\to 0$, so that the infra-red
finiteness of the BFKL amplitude is preserved. Here $v$ is the electroweak
vacuum expectation value parameter, $v^2= 1/(G_F \sqrt{2})$, and
\[
 \mathcal{F}(\vec q^{\: 2}) = 
  4 \int_0^1 dy \int_0^{1 - y} dx
  \frac{1 - 4 x y}
       {1 - \left( \frac{M_{H,\perp}^2}
                        {M_t^2} \right) x y 
        + \left( \frac{\vec q^{\: 2}}
                      {M_t^2} \right) y (1 - y)} \;.
\]
In this way we confirm, up to an irrelevant sign for
$\mathcal{F}(\vec q^{\: 2})$, the results obtained earlier in
Ref.~\cite{DelDuca:2003ba}:
\[
 \mathcal{F}(\vec q^{\: 2})  = \frac{- 4 M_t^2}{M_{H,\perp}^2}
 \left\{ - 2 - \left( \frac{2 \vec q^{\: 2}}{M_{H,\perp}^2} \right)
 \left[ \sqrt{z_1} \, \mathcal{W}(z_1) - \sqrt{z_2} \, \mathcal{W}(z_2) \right]
 \right.
\]
\begin{equation}
 \label{IF_F}
 \left.
 + \frac{1}{2} \left( 1 - \frac{4 M_t^2}{M_{H,\perp}^2} \right) \left[ \mathcal{W}(z_1)^2 - \mathcal{W}(z_2)^2 \right]
 \right\} \; ,
\end{equation}
with $\vec q$ the transverse component of the four-vector $q$, $M_{H,\perp} = \sqrt{M_H^2+|\vec q|^2}$ the Higgs-boson transverse mass, $z_1 = 1
- 4 M_t^2/M_H^2$, $z_2 = 1 + 4 M_t^2/\vec q^{\: 2}$, and the root
$\sqrt{z_1} = i \sqrt{|z_1|}$ is taken for negative values of $z_1$.
Furthermore, we have
\begin{equation}
 \label{IF_W}
 \mathcal{W}(z) = \left\{
 \begin{aligned}
  &- 2 i \arcsin \frac{1}{\sqrt{1 - z}} \; , 
  \qquad &z < 0 \; ; \\ 
  &\ln \frac{1 + \sqrt{z}}{1 - \sqrt{z}} - i \pi \; ,
  \qquad &0 < z < 1 \; ; \\ 
  &\ln \frac{1 + \sqrt{z}}{\sqrt{z} - 1} \; ,
  \qquad &z > 1 \; . 
 \end{aligned}
 \right.
\end{equation}

In the large top-mass limit, our LO impact factor reads
\begin{equation}\label{IF_partonic_mass_limit}
 V^{(0)}_{g\to H}(\vec {q})= \frac{\alpha_s^2}{v^2}\frac{\vec q^{\: 2}}
 {72\pi^2\sqrt{N^2_{c}-1}} \; .
\end{equation}

The inclusion of the gluon PDF allows one to write differential
proton-to-Higgs IF
\begin{equation}\label{IF_hadronic}
  dV^{(0)}_{p\to H}(\vec {q})= \frac{\alpha_s^2}{v^2}\frac{|\mathcal{F}
    (\vec q^{\: 2})|^2}{128\pi^2\sqrt{N^2_{c}-1}}\vec q^{\: 2}dx_H f_{g}(x_H) \; ,
\end{equation}
where the subscript $p$ in the left-hand-side denotes now the proton, $dx_H$
stands for the gluon/Higgs longitudinal momentum fraction.
In order to establish the proper normalization for our impact factor, we insert
into (\ref{IF_hadronic}) a delta function depending on the produced Higgs-boson
transverse momentum $\vec p_{H}$, then the LO result for the impact factor reads 
\begin{equation}\label{IF_hadronic_delta}
 \frac{dV^{(0)}_{p\to H}(\vec {q})}{\vec {q}^{\: 2}}=
 \frac{\alpha_s^2}{v^2}\frac{|\mathcal{F}(\vec q^{\: 2})|^2}{128\pi^2\sqrt{N^2_{c}-1}}\vec q^{\: 2}\int_{0}^{1}dx_H f_{g}(x_H) \frac{d^2\vec p_{H}}{\vec p_{H}^{\: 2}}\delta^{(2)}(\vec p_{H}-\vec {q}) \; .
\end{equation}
For later convenience, we transfer the impact factor to the so called
($\nu,n$)-representation, \emph{i.e.} we express it as superposition of
the eigenfunctions of LO BFKL kernel. The outcome is the following:
\begin{equation}\label{IF_hadronic_projection}
 dV^{(0)}_{p\to H}(\nu,n) = 
 \int d^2\vec {q} \, \frac{dV^{(0)}_{p\to H}(\vec {q})}{\vec {q}^{\: 2}}
 \frac{\big(\vec {q}^{\: 2}\big)^{i\nu-1/2}}{\pi\sqrt2}e^{i n \phi} \; ,
\end{equation}
where $\phi$ is the azimuthal angle of the vector $\vec {q}$.  
Combining Eqs.~(\ref{IF_hadronic_delta}) and~(\ref{IF_hadronic_projection}),
we get the following differential expression for our LO impact factor:
\begin{equation}\label{IF_hadronic_projected}
\frac{dV^{(0)}_{p\to H}(\nu,n)}{dx_Hd^2\vec p_{H}}= \frac{\alpha_s^2}{v^2} 
\frac{|\mathcal{F}(\vec p_{H}^{\: 2})|^2}{128\pi^{3}\sqrt{2(N^2_{c}-1)}} \big(\vec 
 p_{H}^{\: 2}\big)^{i\nu-1/2}f_{g}(x_H) e^{i n \phi_H} \; ,
\end{equation}
where $\phi_H$ denotes the azimuthal angle of the vector $\vec {p_H}$. 

For the sake of completeness, we give the corresponding expression for
the jet LO impact factor~\cite{Ivanov:2012ms}
\begin{equation}\label{IF_jet_projected}
  \frac{d\Phi^{(0)}_{J}(\nu,n)}{dx_J d^2\vec p_J}= 2\alpha_s\sqrt{\frac{C_{F}}
    {C_{A}}}(\vec p_J^{\: 2})^{i\nu-3/2} \bigg(\frac{C_{A}}{C_{F}}f_{g}(x_J)
  +\sum_{a=q\overline{q}}f_{a}(x_J)\bigg) e^{i n \phi_J} \; ,
\end{equation}
where $\phi_J$ denotes the azimuthal angle of the vector $\vec {p_J}$.

In the next section we will build the cross section for the process of our
consideration, by combining the BFKL Green's function and impact factor for
the jet, together with our calculated Higgs-gluon impact factor.

\subsection{Cross section and azimuthal coefficients} 
\label{cross_section}

For the sake of simplicity, we consider final-state configurations where the
Higgs is always tagged in a more forward direction with respect to the jet,
thus implying $\Delta Y \equiv y_H - y_J > 0$.

As anticipated, the Higgs and the jet are also expected to feature large
transverse momenta,
$|\vec p_H|^2 \sim |\vec p_J|^2 \gg \Lambda^2_{\rm QCD}$.
The four-momenta of the parent protons, $p_{1,2}$, are taken as Sudakov vectors
satisfying $p^2_{1,2} = 0$ and $p_1 p_2 = s/2$, so that the final-state
transverse momenta can be decomposed in the following way:
\begin{eqnarray}
p_H &=& x_H p_1 + \frac{M_{H,\perp}^{\, 2}}{x_H s}p_2 + p_{H\perp} \ , \quad
p_{H\perp}^{\, 2} = - |\vec p_H|^2 \ , \nonumber \\
p_J &=& x_J p_2 + \frac{|\vec p_J|^2}{x_J s}p_1 + p_{J\perp} \ , \quad
p_{J\perp}^{\, 2} = - |\vec p_J|^2 \ ,
\label{sudakov}
\end{eqnarray}
with the space part of the four-vector $p_{1\parallel}$ being taken positive;
$M_{H,\perp} = \sqrt{M_H^2+|\vec p_H|^2}$ is the Higgs-boson transverse mass.

The longitudinal-momentum fractions, $x_{H,J}$, for the Higgs and jet are
related to the corresponding rapidities in the center-of-mass frame via the
relations
\begin{equation}\label{rapidities}
 y_H = \frac{1}{2}\ln\frac{x_H^2 s}{M_{H, \perp}^2} 
 \; , \qquad 
 y_J = \frac{1}{2}\ln\frac{|\vec p_J|^2}{x_J^2 s} 
 \; , \qquad
 dy_{H,J} = \pm \frac{dx_{H,J}}{x_{H,J}}
 \; .
\end{equation}
As for the rapidity distance, one has
\begin{equation}\label{rapidity_interval}
 \Delta Y = y_H - y_J = \ln \frac{x_H x_J s}{M_{H,\perp} |\vec p_J|} \;.
\end{equation}

Using QCD collinear factorization to build the (differential) hadronic the
cross section, one has 
\begin{equation}
 \label{dsigma_QCD}
 \frac{d\sigma}{dx_H dx_J d^2\vec p_H d^2\vec p_J}
 =\sum_{i, j = q,{\bar q},g}
 \int_0^1 dx_1 \int_0^1 dx_2 \ 
 f_i \left( x_1, \mu_{F_1} \right) \
 f_j \left(x_2, \mu_{F_2} \right)
 \frac{d{\hat\sigma}_{i,j}
 \left( \hat s, \mu_{F_{1,2}} \right)}
 {dx_H dx_J d^2\vec p_H d^2\vec p_J}\;,
\end{equation}
where the $i, j$ indices run over the parton kinds 
(quarks $q = u, d, s, c, b$; antiquarks $\bar q = \bar u, \bar d, \bar s,
\bar c, \bar b$;  or gluon $g$), $f_{i, j} \left(x, \mu_{F_{1,2}} \right)$ are
the incoming-proton PDFs; 
$x_{1, 2}$ denote the longitudinal fractions of the partons involved in the
hard subprocess, whereas $\mu_{F_{1,2}}$ stand for the factorization scales
characteristic of the two fragmentation regions of the incoming hadrons;
$d\hat\sigma_{i, j}\left(\hat s \right)$ is
the partonic cross section, with $\hat s \equiv x_1 x_2s$ the squared
center-of-mass energy of the parton-parton collision subreaction.
In the present case, the sum over the parton kinds $i$ restricts to the
gluon contribution only, consistently with a LO treatment of the Higgs impact
factor, as discussed in the previous section.

The BFKL cross section can be presented 
(see Ref.~\cite{Caporale:2012ih} for the derivation)
as the Fourier series of the so-called \emph{azimuthal coefficients},
${\cal C}_n$
\begin{equation}
 \frac{d\sigma}
 {dy_H dy_J\, d|\vec p_H| \, d|\vec p_J|d\varphi_H d\varphi_J}
 =\frac{1}{(2\pi)^2}\left[{\cal C}_0 + \sum_{n=1}^\infty  2\cos (n\varphi )\,
 {\cal C}_n\right]\; ,
\end{equation}
where $\varphi=\varphi_H-\varphi_J-\pi$, with $\varphi_{H,J}$ the Higgs and the
jet azimuthal angles. A comprehensive formula for the $\varphi$-averaged
cross section, ${\cal C}_0$, and the other coefficients, ${\cal C}_{n > 0}$,
reads 
\begin{equation}\nonumber
 {\cal C}_n \equiv \int_0^{2\pi} d\varphi_H \int_0^{2\pi} d\varphi_J\,
 \cos(n \varphi) \,
 \frac{d\sigma}{dy_H dy_J \, d|\vec p_H| \, d|\vec p_J| d\varphi_H d\varphi_J}
\end{equation}
\begin{equation}\nonumber
  = \frac{e^{\Delta Y}}{s} \frac{M_{H,\perp}}{|\vec p_H|}
\end{equation}
  \begin{equation}\nonumber
    \times  \int_{-\infty}^{+\infty} d\nu \, \left(\frac{x_J x_H  s}{s_0}
 \right)^{\bar \alpha_s(\mu_{R_c})\left\{\chi(n,\nu)+\bar\alpha_s(\mu_{R_c})
 \left[\bar\chi(n,\nu)+\frac{\beta_0}{8 N_c}\chi(n,\nu)\left[-\chi(n,\nu)
 +\frac{10}{3}+4\ln\left(\frac{\mu_{R_c}}{\sqrt{\vec p_H\vec p_J}}\right)\right]\right]\right\}}
\end{equation}
\begin{equation}\nonumber
 \times \left\{ \alpha_s^2(\mu_{R_1}) c_H(n,\nu,|\vec p_H|, x_H) \right\}
 \left\{ \alpha_s(\mu_{R_2}) [c_J(n,\nu,|\vec p_J|,x_J)]^* \right\} \,
\end{equation}
\begin{equation}\label{Cn_start_WIP}
 \times \left\{1
 + \alpha_s(\mu_{R_1}) \frac{c_H^{(1)}(n,\nu,|\vec p_H|,
 x_H)}{c_H(n,\nu,|\vec p_H|, x_H)}
 + \alpha_s(\mu_{R_2}) \left[\frac{c_J^{(1)}(n,\nu,|\vec p_J|, x_J)}{c_J(n,\nu,|\vec p_J|,
 x_J)}\right]^*  \right\} \; , 
\end{equation}
\\
where $\bar \alpha_s \equiv N_c/\pi \, \alpha_s$, with $N_c$ the QCD color
number,
\begin{equation}
\beta_0=\frac{11}{3} N_c - \frac{2}{3}n_f
\end{equation}
the first coefficient in the expansion of the QCD $\beta$-function ($n_f$ is the active-flavor number),
\begin{equation}
 \chi\left(n,\nu\right) = 2\psi\left(1\right) - \psi\left(\frac{n + 1}{2} + i\nu \right)-\psi\left(\frac{n + 1}{2} - i\nu \right)
\end{equation}
the eigenvalue of the LO BFKL kernel, $c_{H,J}(n,\nu)$ are the Higgs and the jet
LO impact factors in the ($\nu,n$)-space, given by
\begin{equation}
 \label{cH}
 c_H(n,\nu, |\vec p_H|, x_H) = \frac{1}{v^2} \frac{|\mathcal{F}(\vec p_H^{\: 2})|^2}
 {128\pi^{3}\sqrt{2(N^2_{c}-1)}}
 \left( \vec p_H^{\: 2} \right)^ {i\nu + 1/2} f_g(x_H,\mu_{F_1}) \; ,
\end{equation}
\begin{equation}
 \label{cJ}
c_J(n,\nu, |\vec p_{J}|, x_J) = 
2 \sqrt{\frac{C_F}{C_A}}
\left( \vec p_J^{\: 2} \right)^{i\nu-1/2} 
\left( \frac{C_A}{C_F} f_g(x_J,\mu_{F_2}) + \sum_{a = q,\bar{q}}f_a(x_J,\mu_{F_2})
\right) \;. 
\end{equation}
The energy-scale parameter, $s_0$, is arbitrary within NLA accuracy and will
be fixed in our analysis at $s_0 = M_{H,\perp} |\vec p_J|$.
The remaining quantities are the NLO impact-factor corrections,
$c_{H,J}^{(1)}(n,\nu,|\vec p_{H,J}|, x_{H,J})$. The expression for the Higgs NLO
impact factor has not been yet calculated. It is possible, however, 
to include some ``universal'' NLO contributions to the Higgs impact factor,
which can be expressed through the corresponding LO impact factor, and 
are fixed by the requirement of stability within the NLO under variations
of the energy scale $s_0$, the renormalization scale $\mu_R$ and of the
factorization scale $\mu_F$, getting
\begin{equation}
  \alpha_s c_H^{(1)}(n,\nu,|\vec p_H|, x_H) \to \bar \alpha_s \tilde
  c_H^{(1)}(n,\nu,|\vec p_H|, x_H) \; ,
\end{equation}
with
\[
 \tilde c_H^{(1)}(n,\nu,|\vec p_H|, x_H) =
 c_H(n, \nu, |\vec p_H|, x_H) \left\{ \frac{\beta_0}{4 N_c} 
 \left( 2 \ln \frac{\mu_{R_1}}{|\vec p_H|} + \frac{5}{3} \right)
 + \frac{\chi\left(n,\nu\right)}{2} \ln \left( \frac{s_0}{M_{H,\perp}^2} \right)
 \right.
\]
\begin{equation}
 \label{cH1}
+ \frac{\beta_0}{4 N_c} \left( 2 \ln \frac{\mu_{R_1}}{M_{H,\perp}}\right)
\end{equation}
\[
\left.
-\frac{1}{2N_c f_g(x_H,\mu_{F_1})} \ln \frac{\mu_{F_1}^2}{M_{H,\perp}^2} \int_{x_H}^1\frac{dz}{z}
\left[P_{gg}(z)f_g\left(\frac{x_H}{z},\mu_{F_1}\right)+\sum_{a={q,\bar q}} P_{ga}(z)
    f_a\left(\frac{x_H}{z},\mu_{F_1}\right)\right]
\right\} \;.
\]

The jet impact factor is known at the
NLO~\cite{bar1,bar2,Caporale:2011cc,Ivanov:2012ms,Colferai:2015zfa},
and at that order it is not universal, since it depends on the
  adopted jet selection function (see, {\it e.g.},
  Ref.~\cite{Colferai:2015zfa}). Nonetheless
we treated it on the same ground as the Higgs one, including only the NLO
corrections fixed by the renormalization group and leading to  
\begin{equation}
\label{cJ1}
\tilde c_J^{(1)}(n,\nu,|\vec p_J|, x_J) =
c_J(n, \nu, |\vec p_J|, x_J) \left\{ \frac{\beta_0}{4 N_c} 
\left( 2 \ln \frac{\mu_{R_2}}{|\vec p_J|} + \frac{5}{3} \right)
+ \frac{\chi\left(n,\nu\right)}{2} \ln \left( \frac{s_0}{|\vec p_J|^2} \right)
\right.
\end{equation}
\[
-\frac{1}{2N_c \left( \frac{C_A}{C_F} f_{g}(x_J,\mu_{F_2}) +
  \sum_{a = q,\bar{q}}f_{a}(x_J,\mu_{F_2}) \right) }
\ln \frac{\mu_{F_2}^2}{|\vec p_J|^2}
\]
\[
\times \biggl(
\frac{C_A}{C_F} 
\int_{x_J}^1\frac{dz}{z}
\left[P_{gg}(z)f_g\left(\frac{x_J}{z},\mu_{F_2}\right)+\sum_{a={q,\bar q}} P_{ga}(z)
    f_a\left(\frac{x_J}{z},\mu_{F_2}\right)\right]\biggr.
\]
\[
\left.\biggl.
+ \sum_{a = q,\bar{q}} \int_{x_J}^1\frac{dz}{z}
\left[P_{ag}(z) f_g\left(\frac{x_J}{z},\mu_{F_2}\right)
  + P_{aa}(z)f_a\left(\frac{x_J}{z},\mu_{F_2}\right)\right] \biggr)
\right\} \; .
\]

Combining all the ingredients, we can write our master formula for the
azimuthal coefficients,
\begin{equation}\nonumber
 {\cal C}_n = \frac{e^{\Delta Y}}{s} \frac{M_{H,\perp}}{|\vec p_H|}
\end{equation}
\begin{equation}\nonumber
\times  \int_{-\infty}^{+\infty} \!\!\! d\nu \left(\frac{x_J x_H s}{s_0}
 \right)^{\bar \alpha_s(\mu_{R_c})\left\{\chi(n,\nu)+\bar\alpha_s(\mu_{R_c})
 \left[\bar\chi(n,\nu)+\frac{\beta_0}{8 N_c}\chi(n,\nu)\left[-\chi(n,\nu)
 +\frac{10}{3}+4\ln\left(\frac{\mu_{R_c}}{\sqrt{\vec p_H\vec p_J}}\right)\right]\right]\right\}}
\end{equation}
\begin{equation}
 \times \left\{ \alpha_s^2(\mu_{R_1}) c_H(n,\nu,|\vec p_H|, x_H) \right\}
 \left\{ \alpha_s(\mu_{R_2}) [c_J(n,\nu,|\vec p_J|,x_J)]^* \right\} \,
\end{equation}
\begin{equation}\label{Cn_master}\nonumber
 \times \left\{1
 + \bar \alpha_s(\mu_{R_1}) \frac{\tilde c_H^{(1)}(n,\nu,|\vec p_H|,
 x_H)}{c_H(n,\nu,|\vec p_H|, x_H)}
 + \bar \alpha_s(\mu_{R_2}) \left[\frac{\tilde c_J^{(1)}(n,\nu,|\vec p_J|, x_J)}{c_J(n,\nu,|\vec p_J|,
 x_J)}\right]^*  \right\} \; . 
\end{equation}
The renormalization scales ($\mu_{R_{1,2,c}}$) and the factorization ones
($\mu_{F_{1,2}}$) can, in principle, be chosen arbitrarily, since their variation
  produces effects beyond the NLO. It is however advisable to relate them to the
  physical hard scales of the process.
We chose to fix them differently from each other, depending on the subprocess
to which they are related: $\mu_{R_1} \equiv \mu_{F_1} = C_\mu M_{H,\perp}$,
$\mu_{R_2} \equiv \mu_{F_2} = C_\mu |\vec p_J|$,
$\mu_{R_c} = C_\mu \sqrt{M_{H,\perp} |\vec p_J|}$, where $C_\mu$ is a variation
parameter introduced to gauge the effect of a change of the scale (see the
discussion at the end of Section~\ref{results}).

\section{Phenomenology}
\label{phenomenology}

\subsection{Azimuthal correlations and $p_T$-distribution}
\label{observables}

The first observables of our consideration are the azimuthal-angle coefficients
\emph{integrated} over the phase space for two final-state particles, while
the rapidity interval, $\Delta Y$, between the Higgs boson and the jet is
kept fixed:
\begin{equation}
 \label{Cn_int}
 C_n(\Delta Y, s) =
 \int_{p^{\rm min}_H}^{p^{\rm max}_H}d|\vec p_H|
 \int_{p^{\rm min}_J}^{{p^{\rm max}_J}}d|\vec p_J|
 \int_{y^{\rm min}_H}^{y^{\rm max}_H}dy_H
 \int_{y^{\rm min}_J}^{y^{\rm max}_J}dy_J
 \, \delta \left( y_H - y_J - \Delta Y \right)
 \, {\cal C}_n 
 \, .
\end{equation}
Pursuing the goal of fitting realistic kinematic cuts adopted by the current
experimental analyses at the LHC, we constrain the Higgs emission inside the
rapidity acceptances of the CMS barrel detector, \emph{i.e.} $|y_H| < 2.5$,
while we allow for a larger rapidity range of the
jet~\cite{Khachatryan:2016udy}, which can be detected also by the CMS endcaps,
namely $|y_J| < 4.7$.
Furthermore, three distinct cases for the final-state transverse momenta are
considered:
\begin{itemize}
\item[a)]
  \textit{\textbf{symmetric}} configuration, suited to the search of pure BFKL
  effects, where both the Higgs and the jet transverse momenta lie in the
  range: 20 GeV $< |\vec p_{H,J}| <$ 60 GeV; \,
\item[b)]
  \textit{\textbf{asymmetric}} selection, typical of the ongoing LHC
  phenomenology, where the Higgs transverse momentum runs from 10 GeV to
  $2 M_t$, where the jet is tagged inside its typical CMS configuration, from
  20 to 60 GeV;
\item[c)]
  \textit{\textbf{disjoint windows}}, which allows for the maximum
  exclusiveness in the final state: 35 GeV $< |\vec p_J| <$ 60 GeV and
  60 GeV $< |\vec p_H| < 2 M_t$.
\end{itemize}
We study the $\varphi$-averaged cross section (\emph{alias} the
$\Delta Y$-distribution), $C_0(\Delta Y, s)$, the azimu\-thal-correlation
moments, $R_{n0}(\Delta Y, s) = C_{n}/C_{0} \equiv \langle \cos n \varphi \rangle$,
and their ratios, $R_{nm} = C_{n}/C_{m}$~\cite{Vera:2006un,Vera:2007kn} as
functions of the Higgs-jet rapidity distance, $\Delta Y$.

The second observable of our interest is the $p_H$-distribution for a given
value of $\Delta Y$:
\begin{equation}
 \label{pT_distribution}
 \frac{d\sigma(|\vec p_H|, \Delta Y, s)}{d|\vec p_H| d\Delta Y} =
 \int_{p^{\rm min}_J}^{{p^{\rm max}_J}}d|\vec p_J|
 \int_{y^{\rm min}_H}^{y^{\rm max}_H}dy_H
 \int_{y^{\rm min}_J}^{y^{\rm max}_J}dy_J
 \, \delta \left( y_H - y_J - \Delta Y \right)
 \, {\cal C}_0 
 \, ,
\end{equation}
the Higgs and jet rapidity ranges being given above and 35 GeV $< |\vec p_J| <$
60 GeV.

\subsection{Results and discussion}
\label{results}

\begin{figure}[t]
\centering
\includegraphics[scale=0.53,clip]{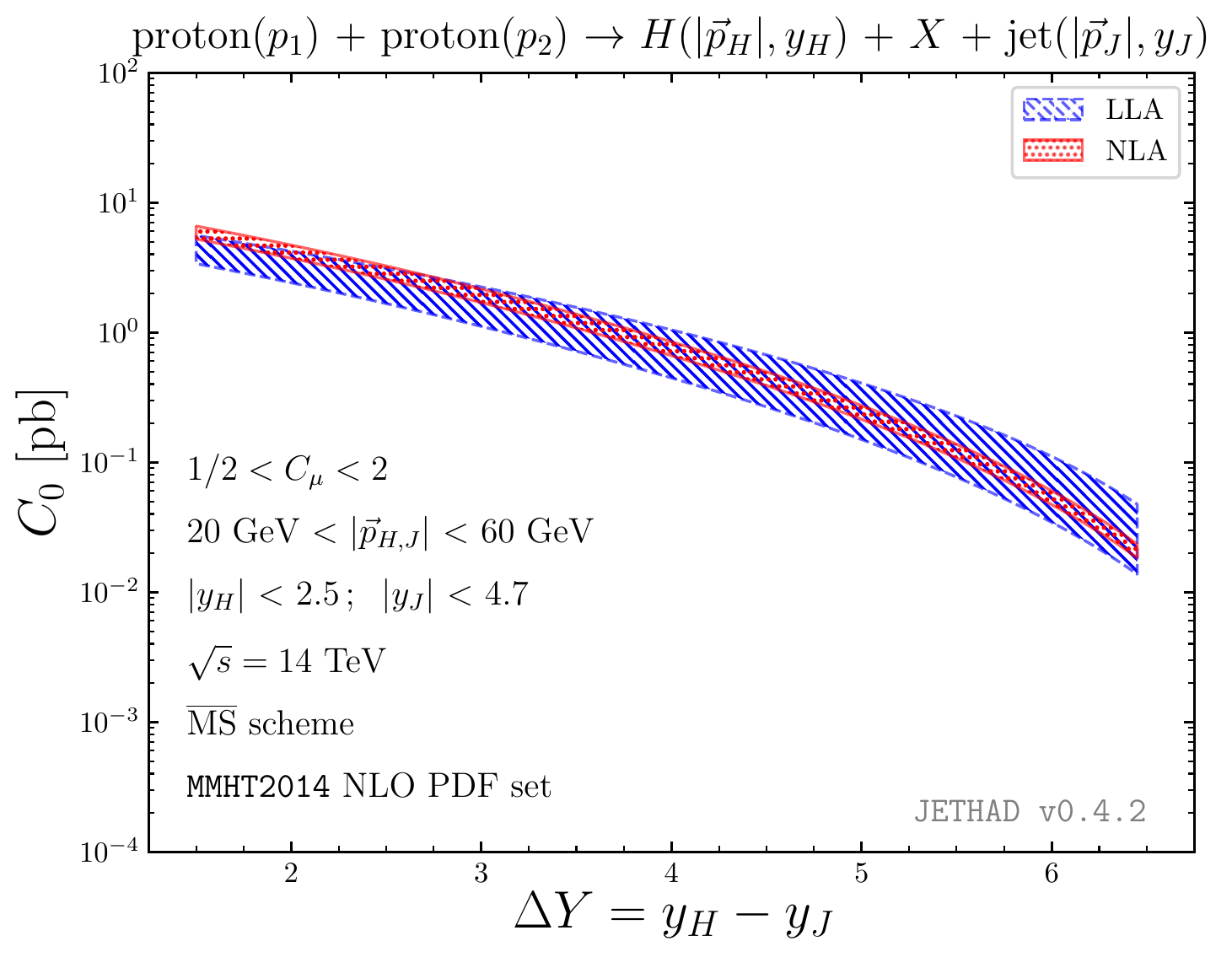}
\includegraphics[scale=0.53,clip]{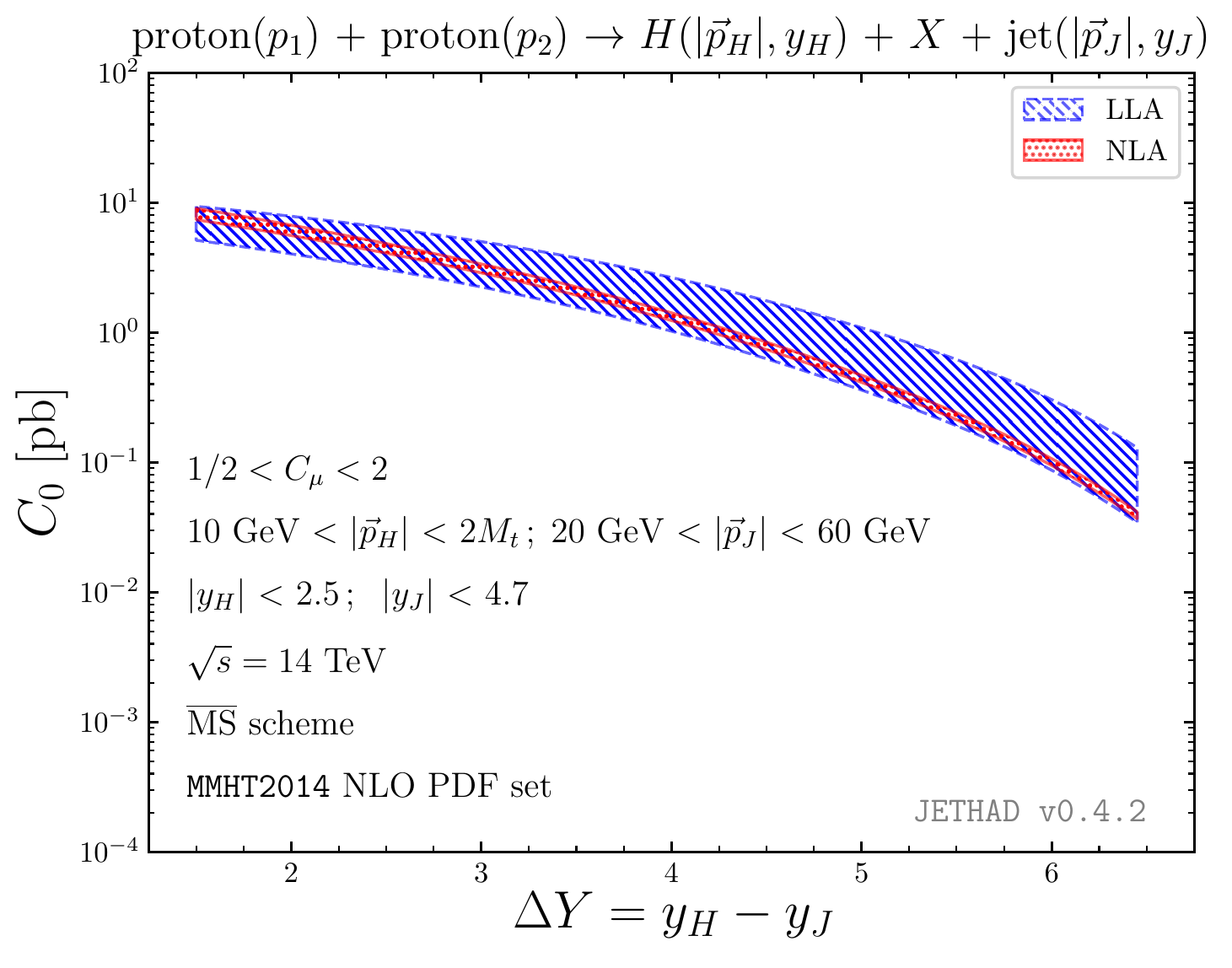}
\includegraphics[scale=0.53,clip]{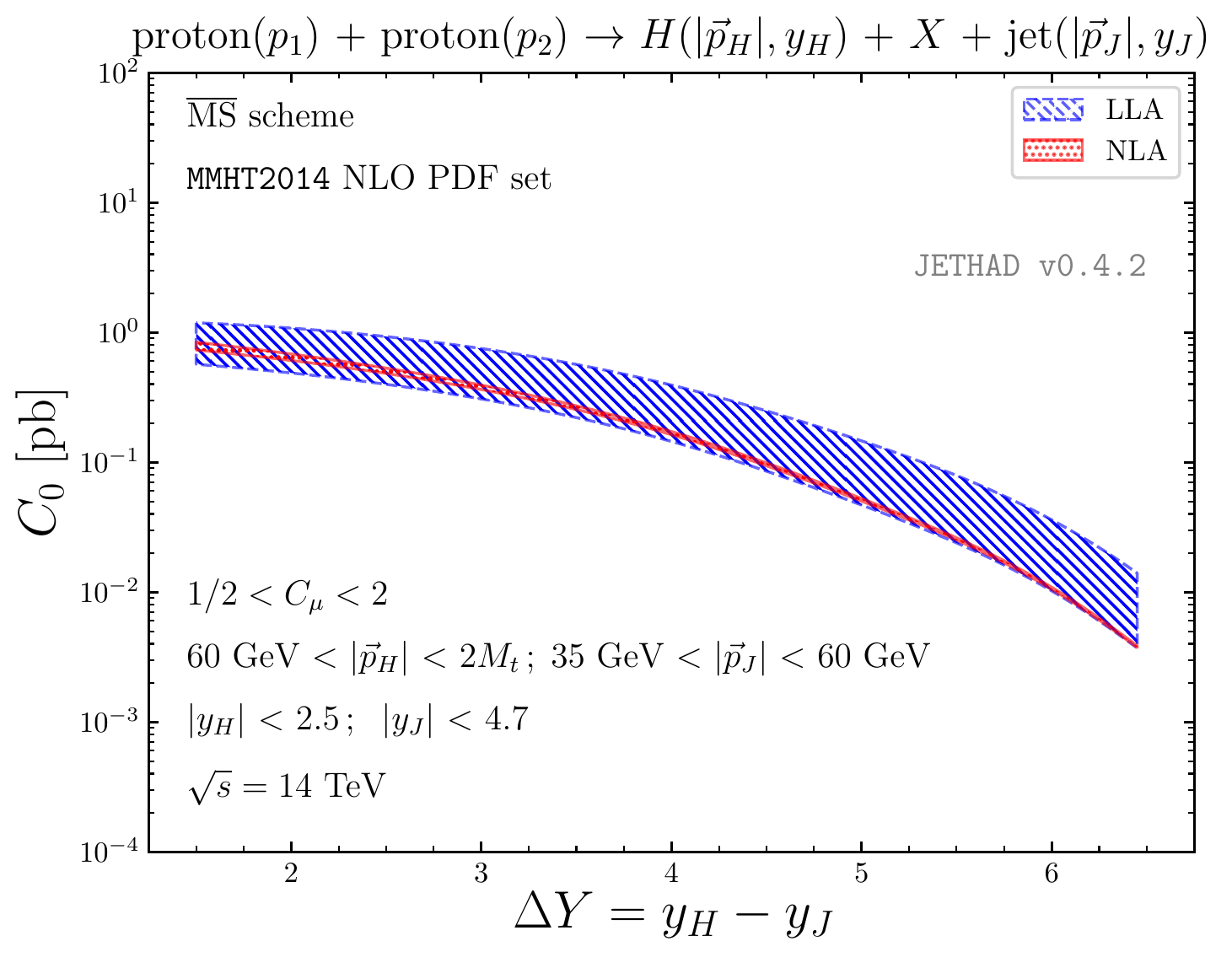}
\caption{$\Delta Y$-dependence of the $\varphi$-averaged cross section, $C_0$, for the inclusive Higgs-jet hadroproduction in the three considered $p_T$-ranges and for $\sqrt{s} = 14$ TeV. Shaded bands give the combined effect of the scale variation with the uncertainty coming from the phase-space numerical integration.}
\label{fig:C0_kt-asw}
\end{figure}

\begin{figure}[p]
\centering
\includegraphics[scale=0.54,clip]{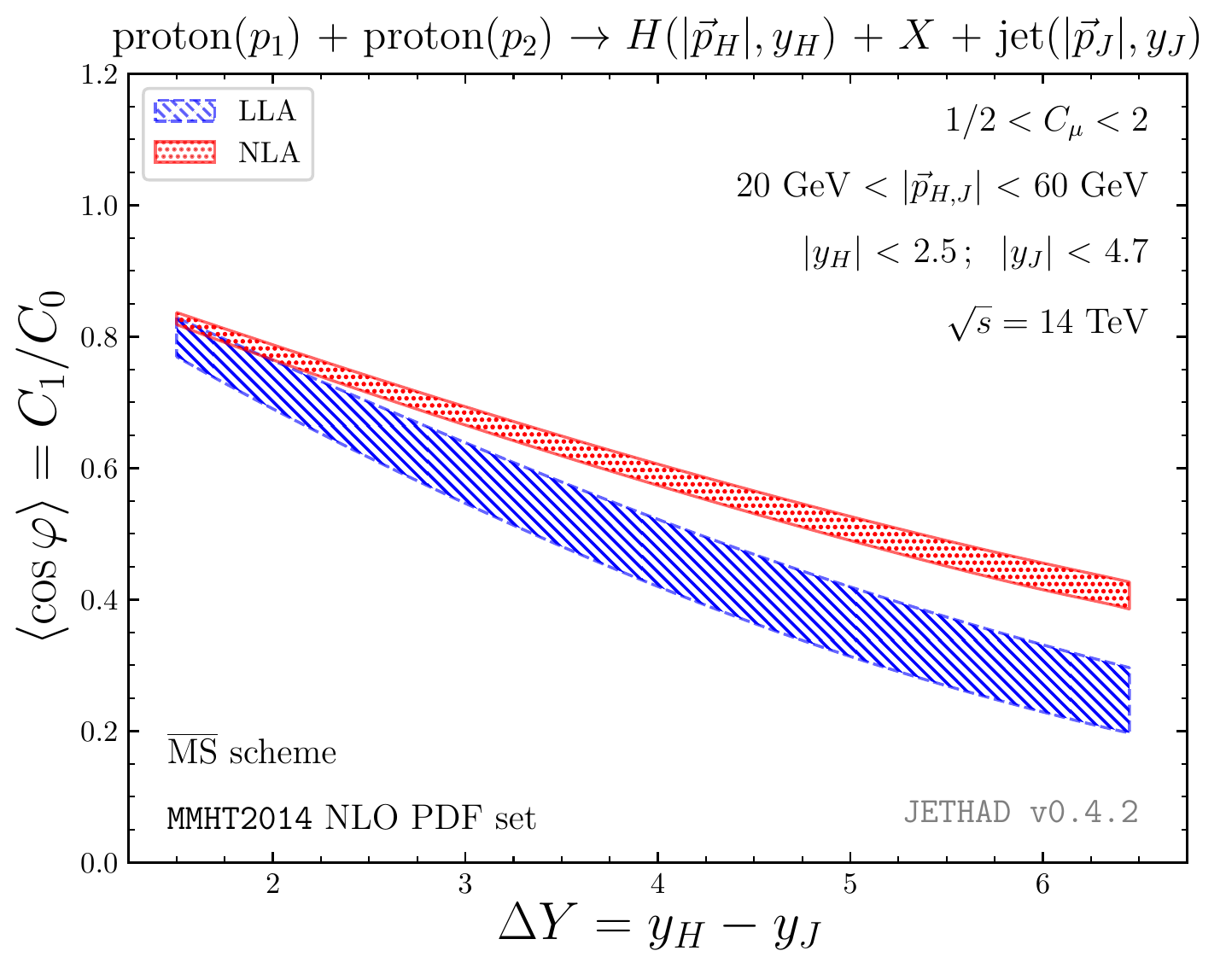}
\includegraphics[scale=0.54,clip]{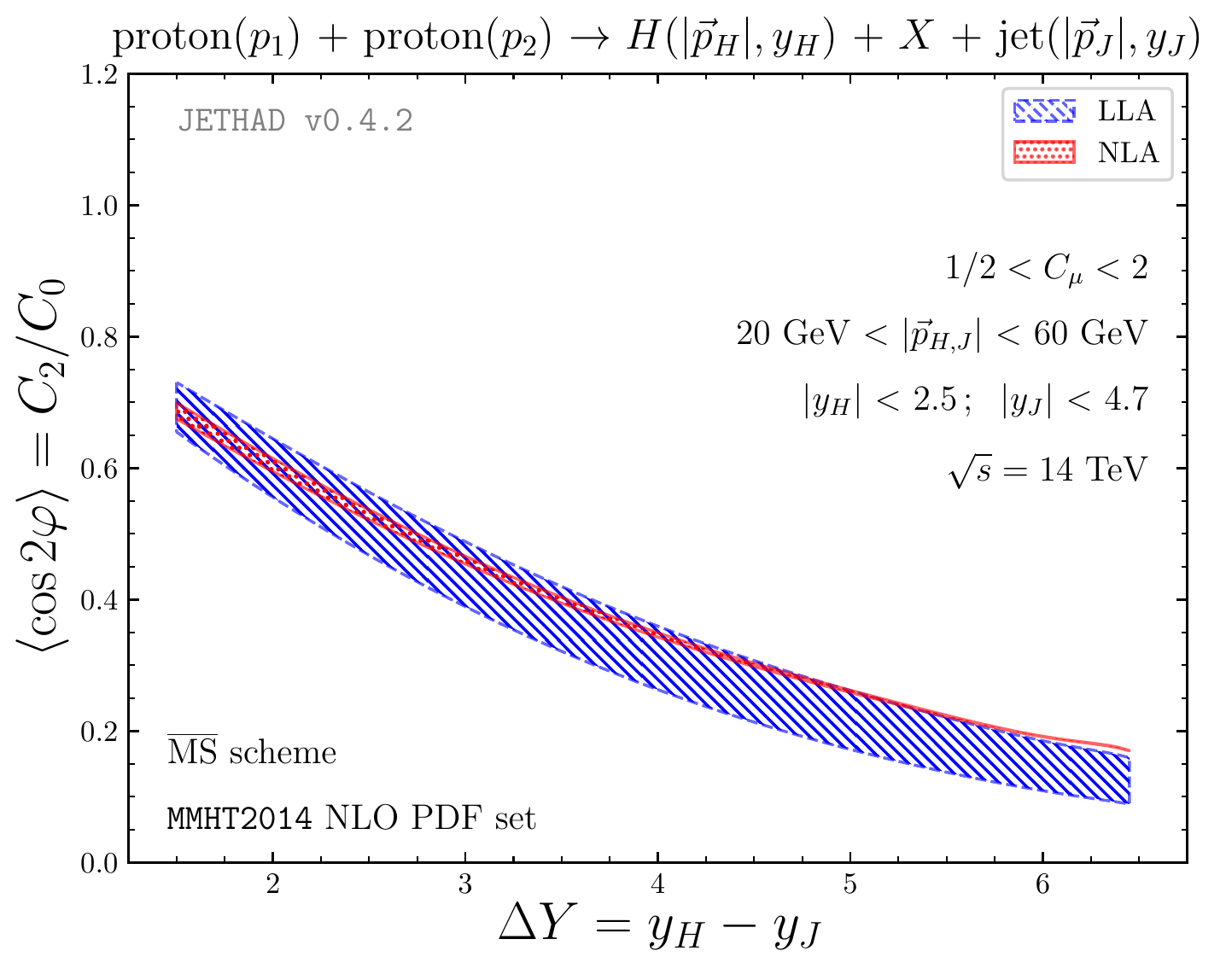}

\includegraphics[scale=0.54,clip]{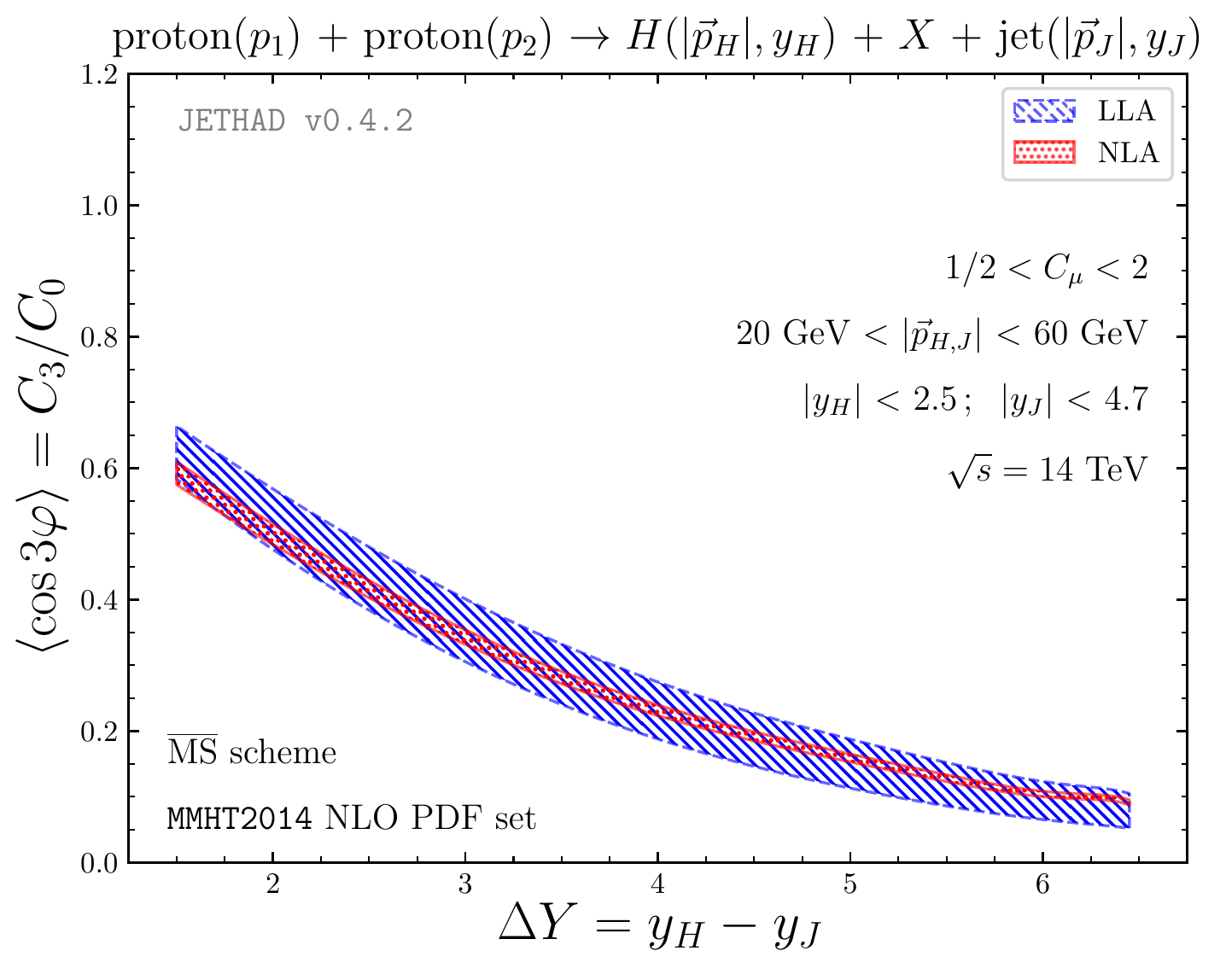}
\includegraphics[scale=0.54,clip]{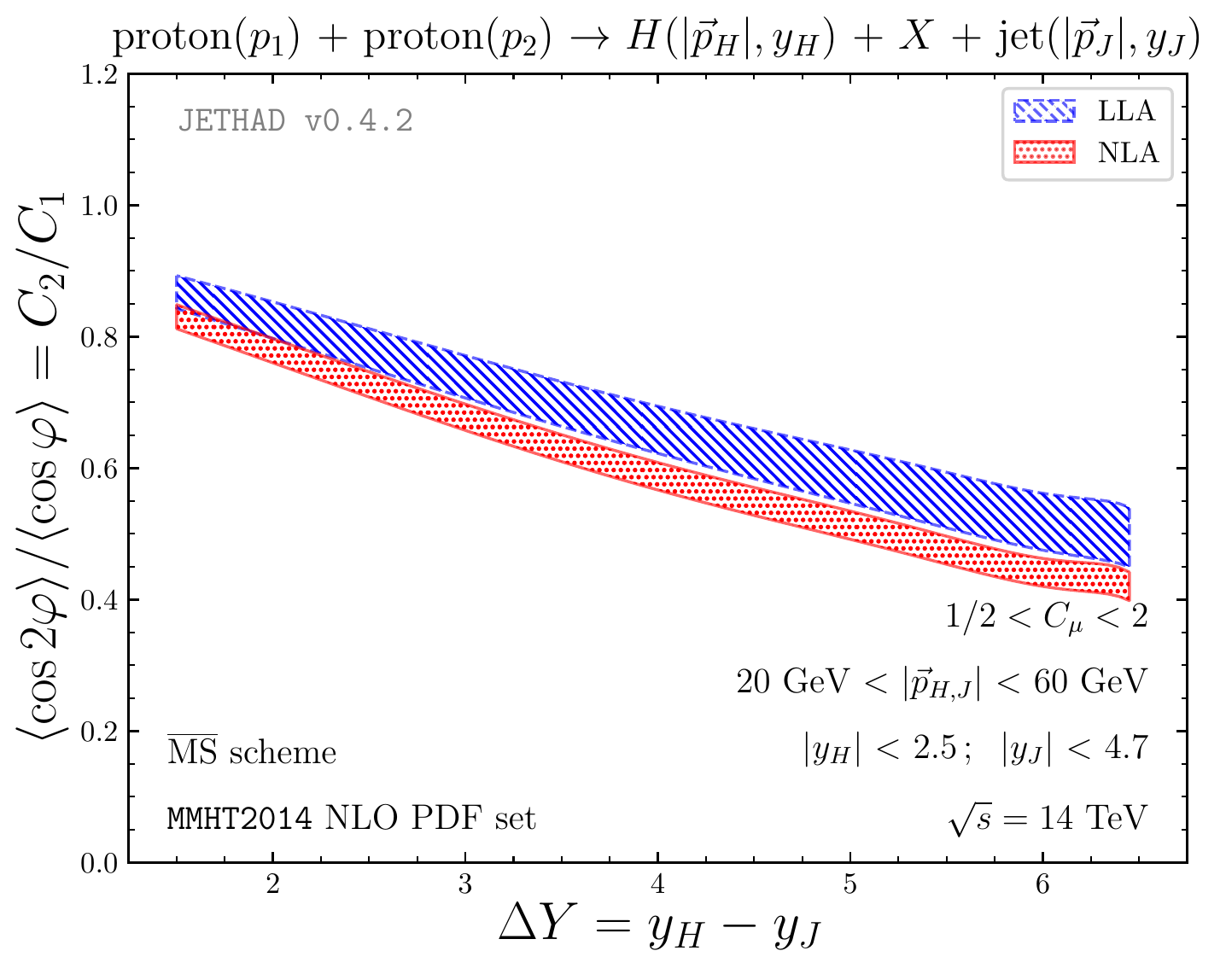}

\includegraphics[scale=0.54,clip]{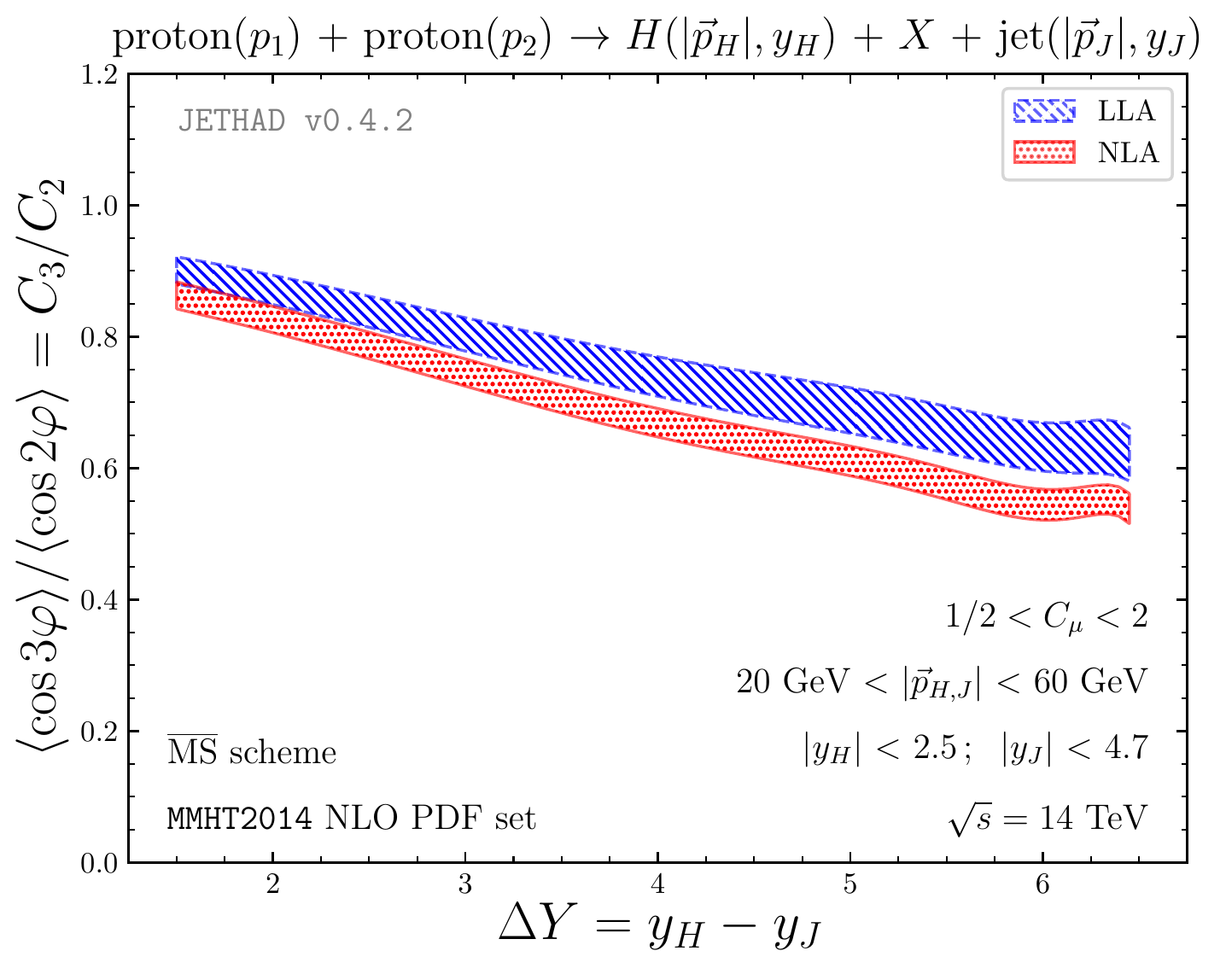}
\caption{$\Delta Y$-dependence of several ratios $R_{nm} \equiv C_{n}/C_{m}$,
  for the inclusive Higgs-jet hadroproduction in the $p_T$-\emph{symmetric}
  configuration and for $\sqrt{s} = 14$ TeV. Shaded bands give the combined
  effect of the scale variation with the uncertainty coming from the phase-space numerical integration.}
\label{fig:Rnm_kt-s}
\end{figure}

\begin{figure}[p]
\centering
\includegraphics[scale=0.54,clip]{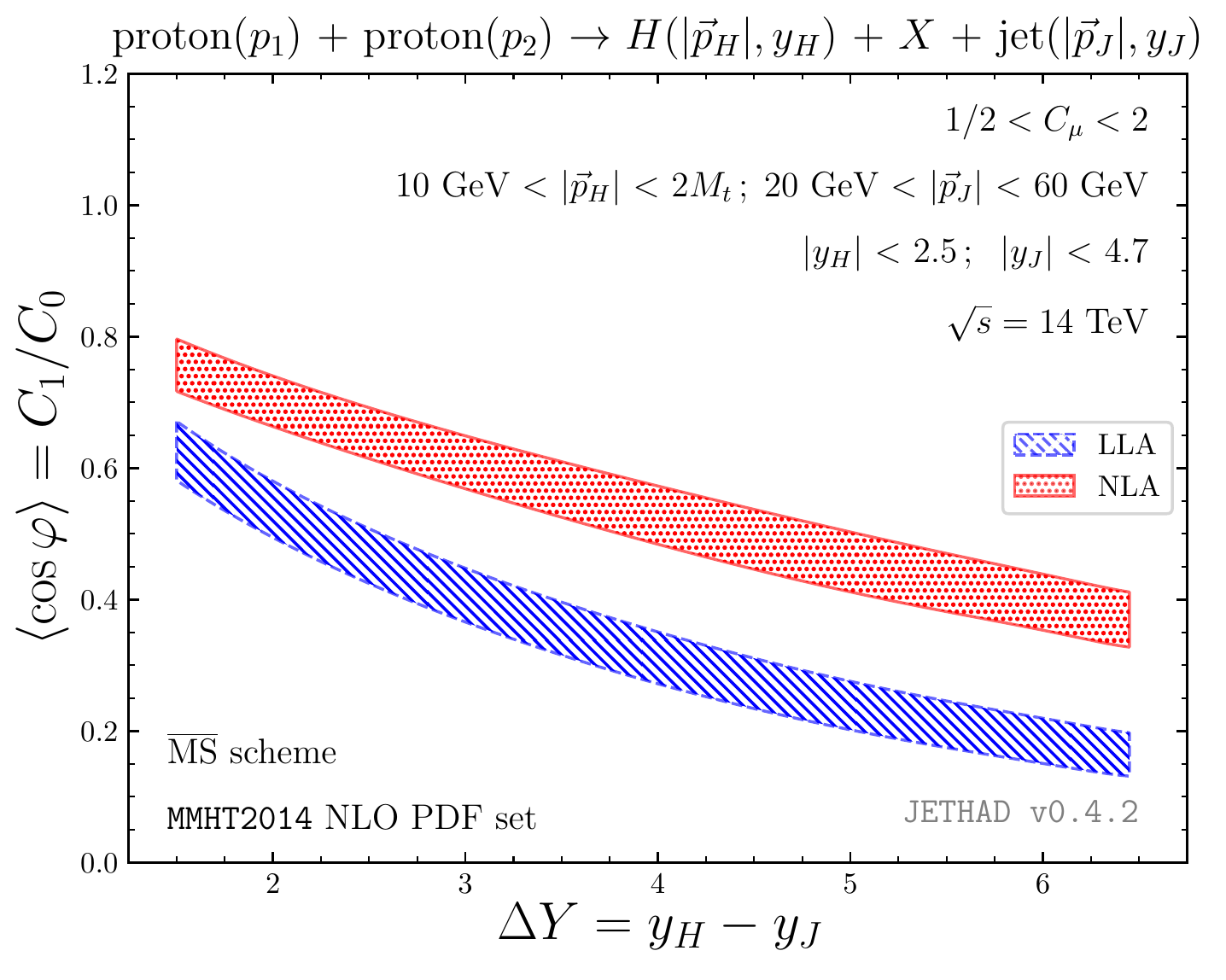}
\includegraphics[scale=0.54,clip]{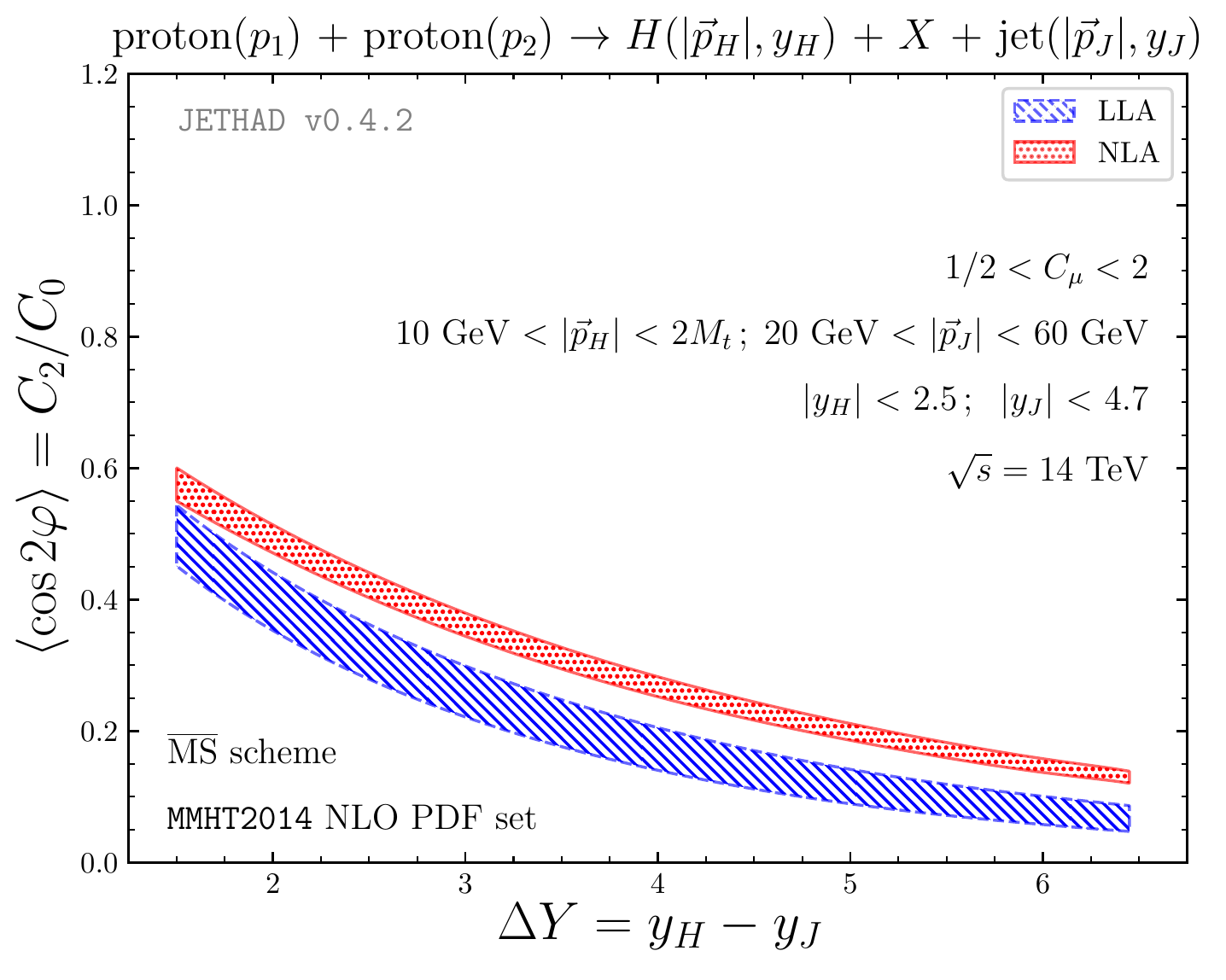}

\includegraphics[scale=0.54,clip]{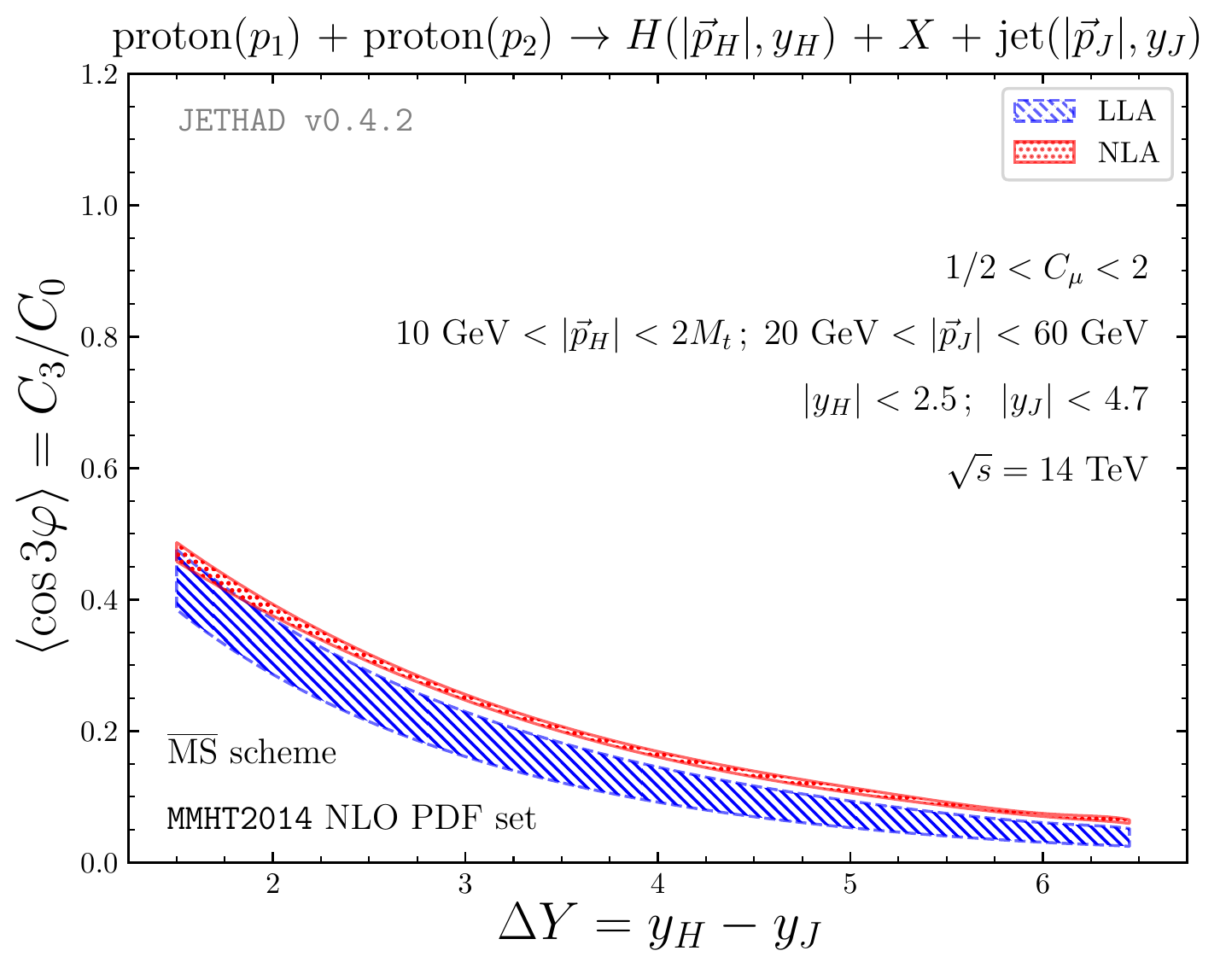}
\includegraphics[scale=0.54,clip]{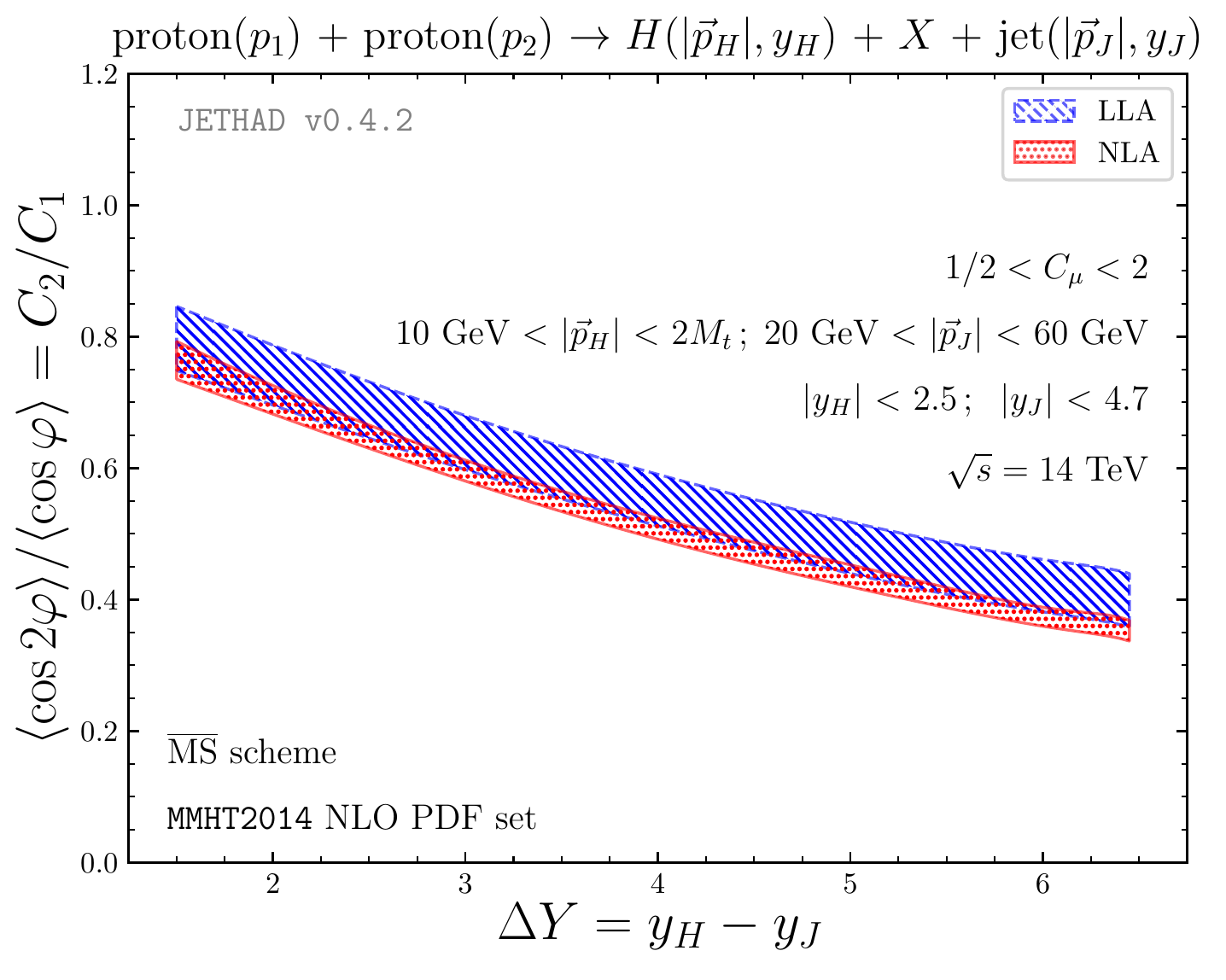}

\includegraphics[scale=0.54,clip]{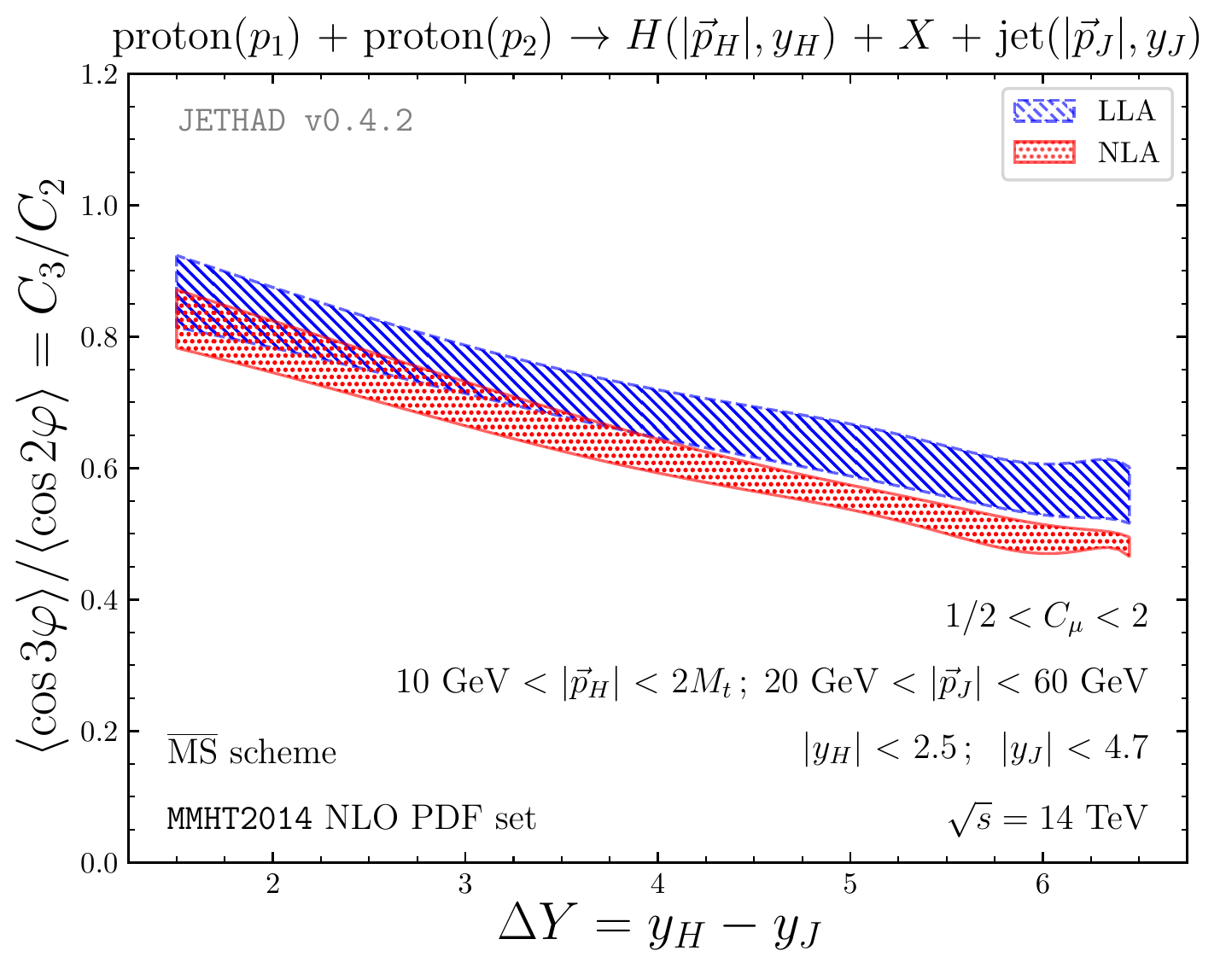}
\caption{$\Delta Y$-dependence of several ratios $R_{nm} \equiv C_{n}/C_{m}$,
  for the inclusive Higgs-jet hadroproduction in the $p_T$-\emph{asymmetric}
  configuration and for $\sqrt{s} = 14$ TeV. Shaded bands give the combined
  effect of the scale variation with the uncertainty coming from the
  phase-space numerical integration.}
\label{fig:Rnm_kt-a}
\end{figure}

\begin{figure}[p]
\centering
\includegraphics[scale=0.54,clip]{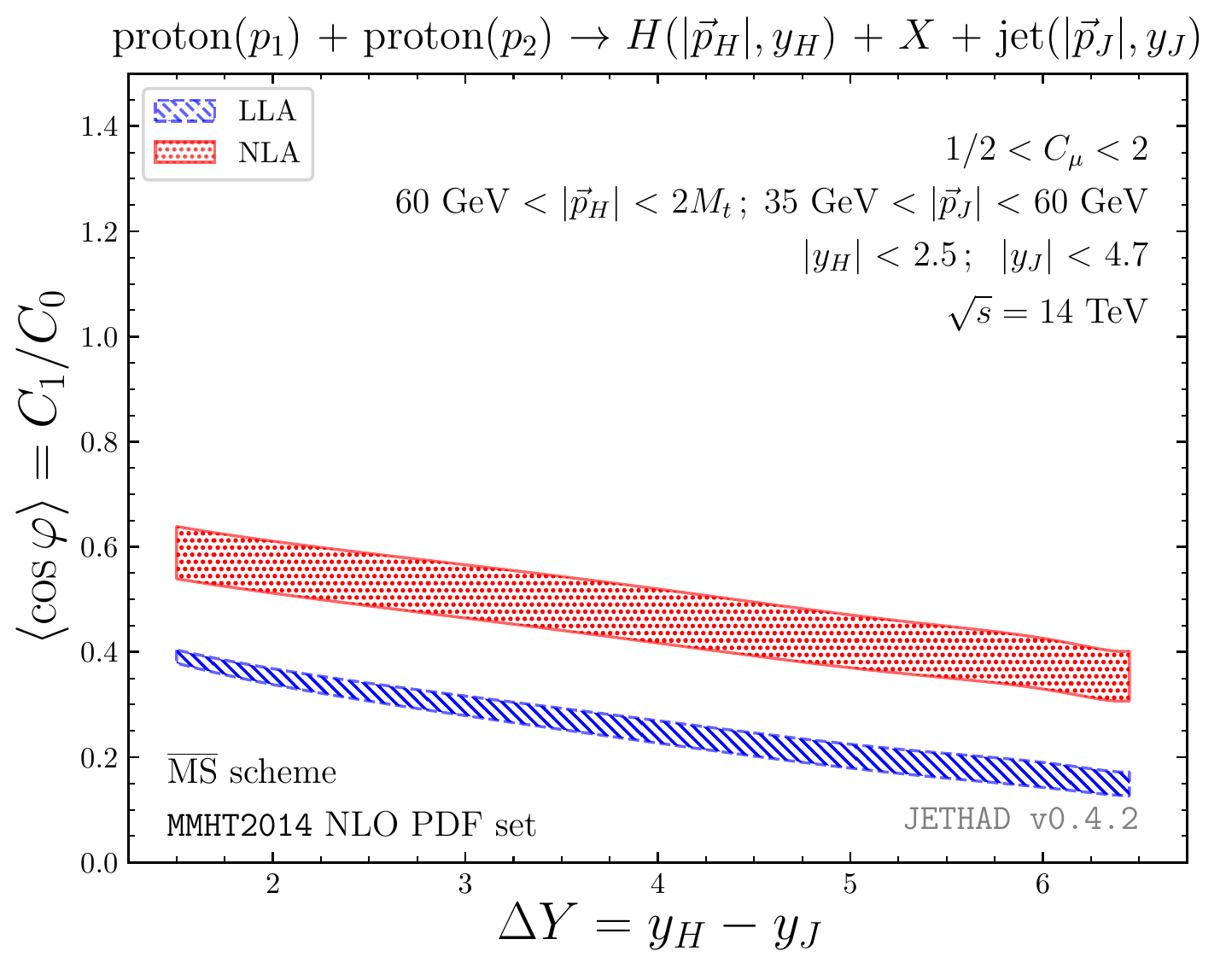}
\includegraphics[scale=0.54,clip]{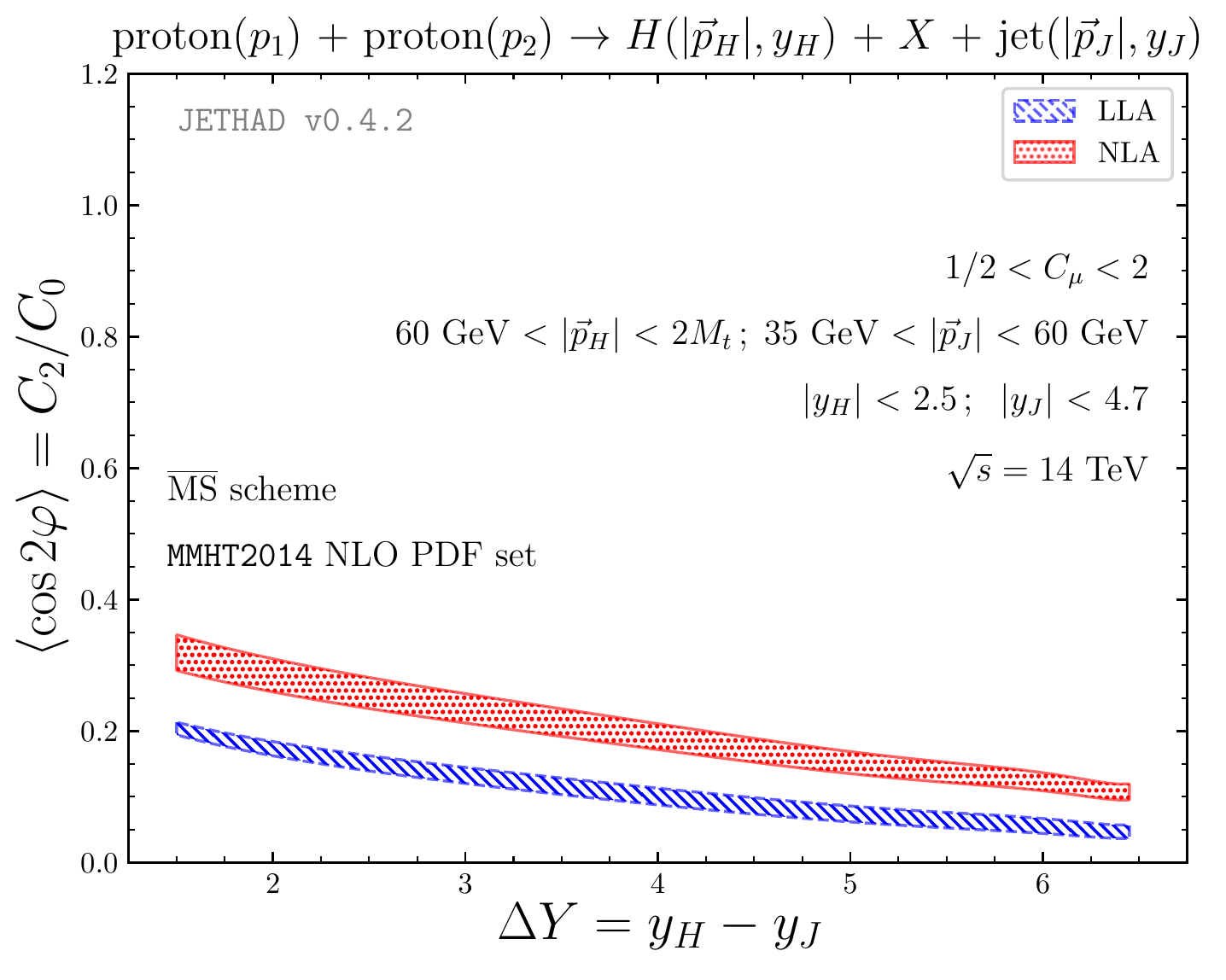}

\includegraphics[scale=0.54,clip]{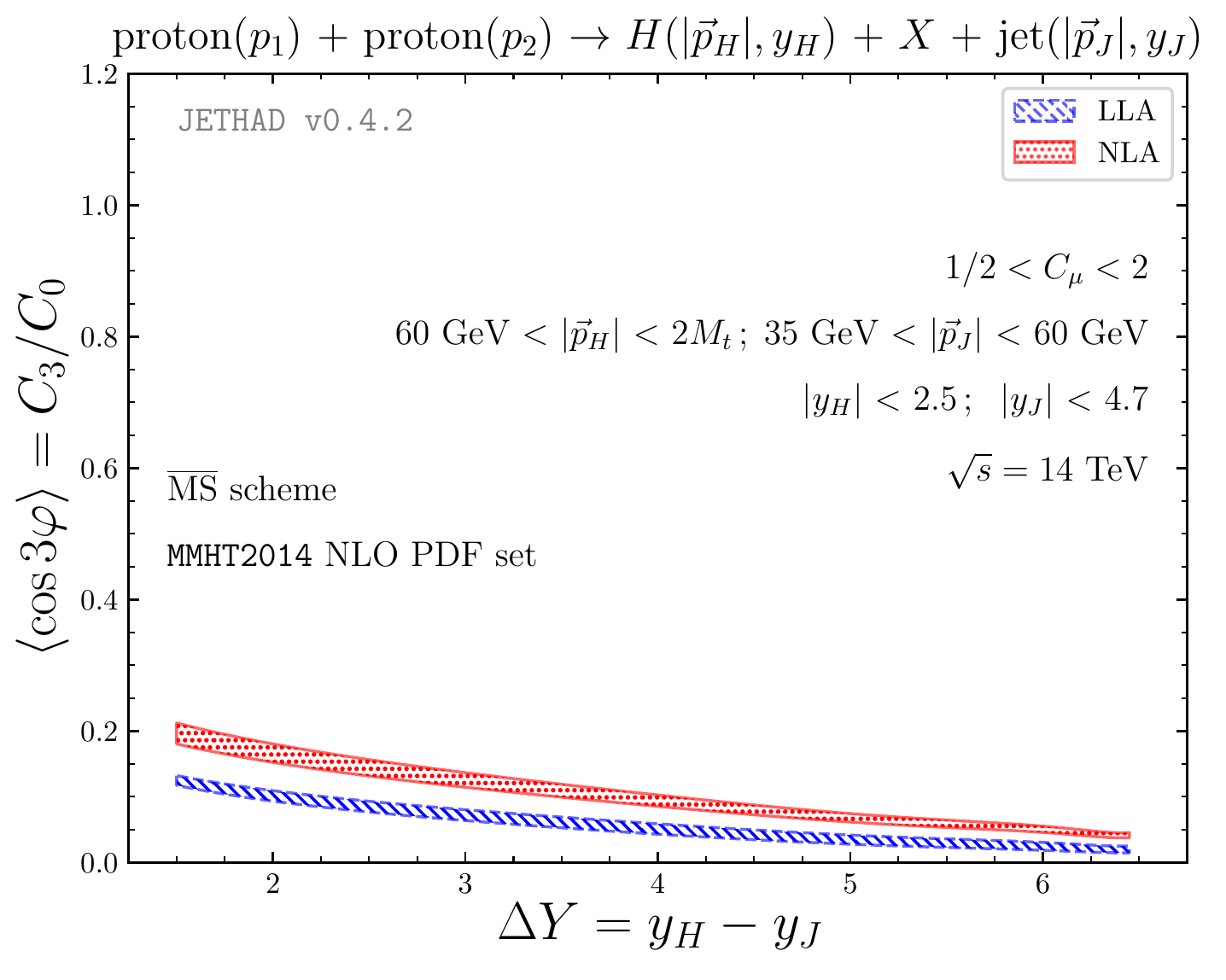}
\includegraphics[scale=0.54,clip]{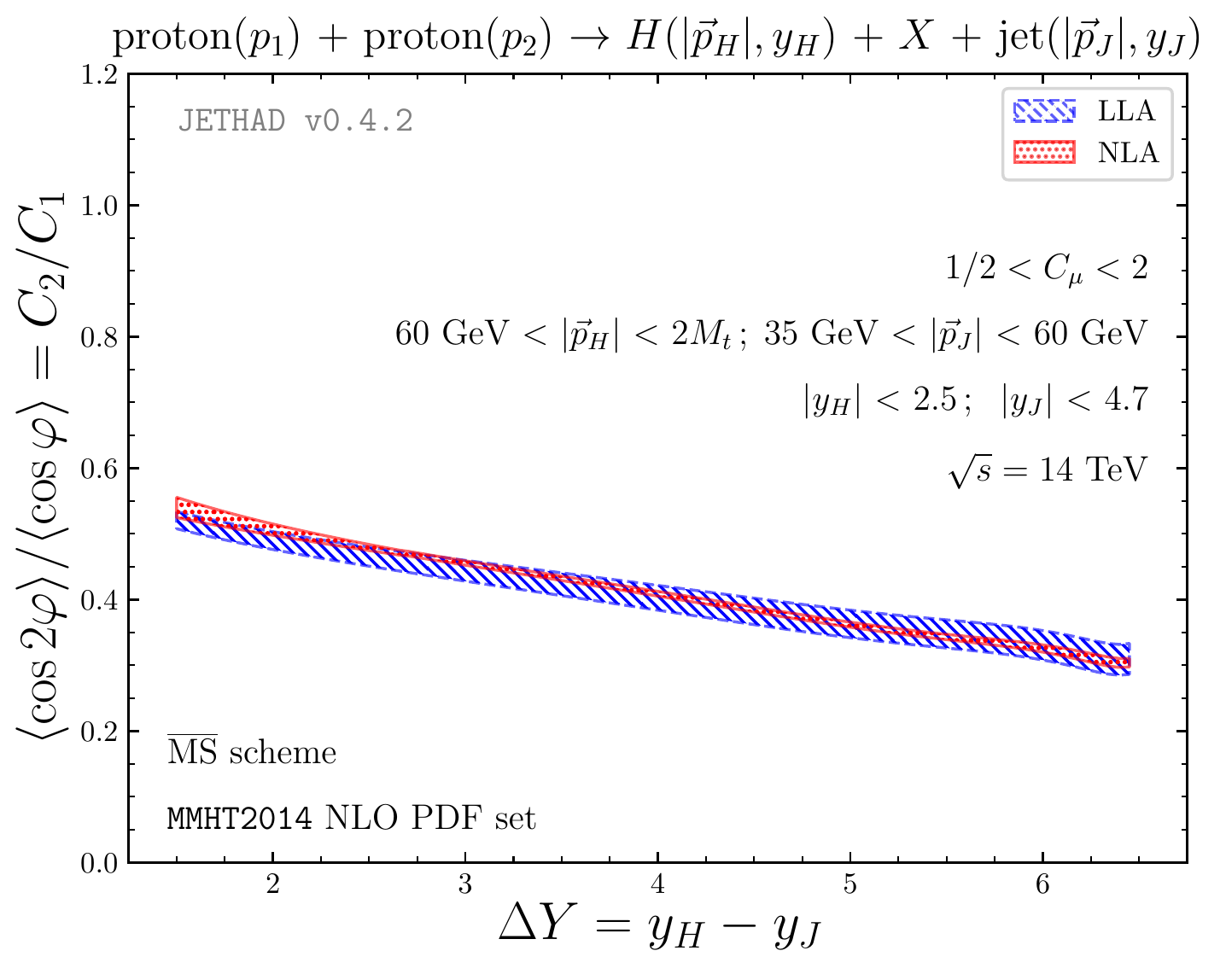}

\includegraphics[scale=0.54,clip]{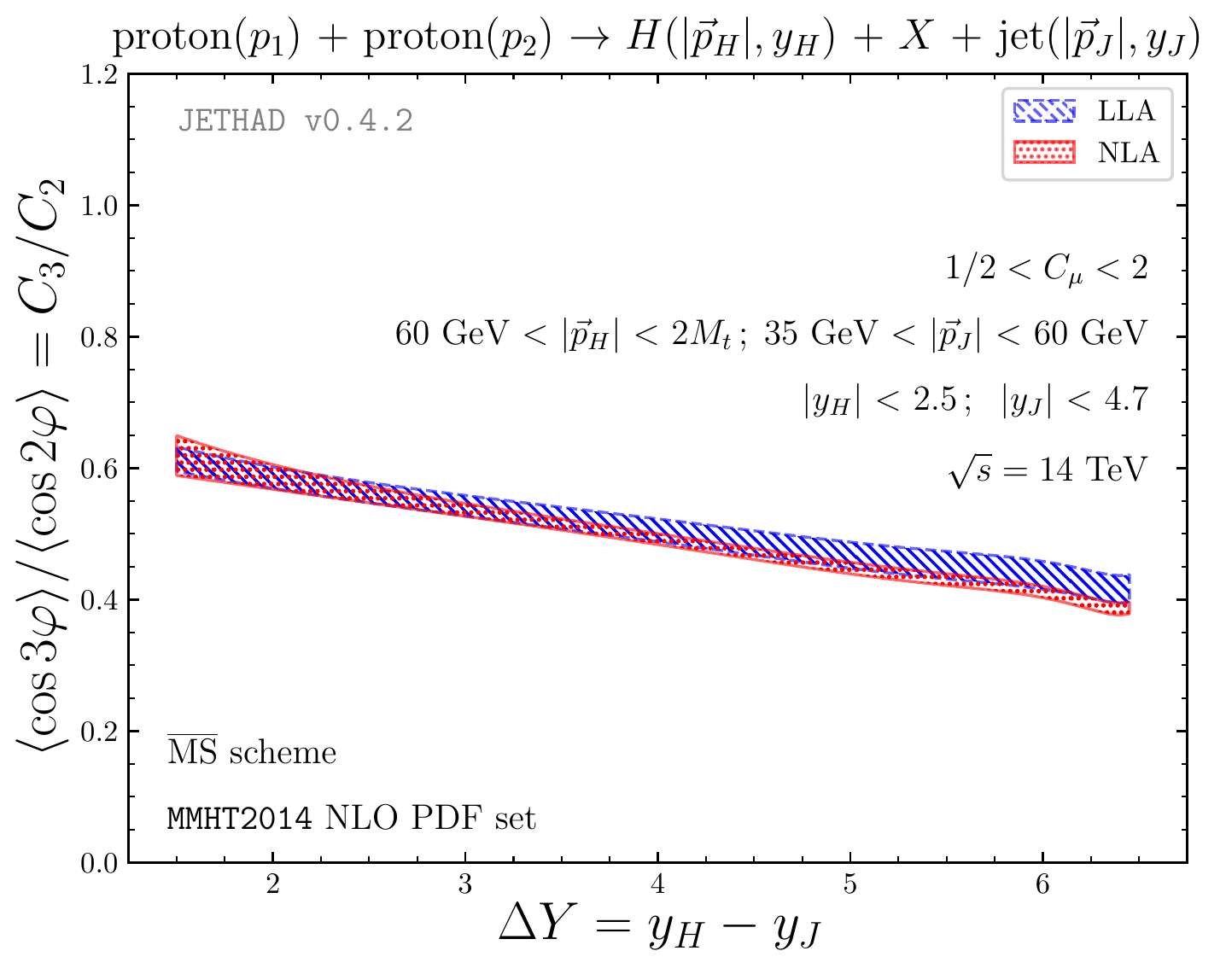}
\caption{$\Delta Y$-dependence of several ratios $R_{nm} \equiv C_{n}/C_{m}$,
  for the inclusive Higgs-jet hadroproduction in the \emph{disjoint}
  $p_T$-windows configuration and for $\sqrt{s} = 14$ TeV. Shaded bands give
  the combined effect of the scale variation with the uncertainty coming from
  the phase-space numerical integration.}
\label{fig:Rnm_kt-w}
\end{figure}

In Fig.~\ref{fig:C0_kt-asw} we present results for the $\Delta Y$-distribution,
$C_0$, in the three kinematic configurations under investigation. 
Here, the usual onset of the BFKL dynamics comes easily out. The growth with
energy of the pure partonic cross sections is quenched by the convolution with
PDFs, this leading to a lowering with $\Delta Y$ of hadronic distributions. 
Notably, NLA predictions (red) are almost entirely contained
  inside LLA uncertainty bands (blue),
thus corroborating the underlying assumption that
the large energy scales provided by the emission of a Higgs boson stabilize
the BFKL series.
A further manifestation of this effect appears in the analysis of azimuthal
correlations, $R_{nm}$. For all the considered cases (Figs.~\ref{fig:Rnm_kt-s},
\ref{fig:Rnm_kt-a} and~\ref{fig:Rnm_kt-w}), higher-order corrections show a
milder discrepancy with respect to pure LLA ones. This represents a novel
feature in the context of semi-hard reactions, where LLA moments have always
shown a fairly stronger decorrelation than NLA ones. Previous studies of
Mueller-Navelet jet production~\cite{Ducloue:2013bva,Caporale:2014gpa,Caporale:2015uva} have highlighted how the use of scale-optimization procedures is
needed to bring NLA patterns near LLA ones and, ultimately, to match CMS
data~\cite{Khachatryan:2016udy}. Conversely, Higgs-jet hadroproduction
genuinely exhibits a solid stability under higher-order corrections in the
range between 1/2 and two times the \emph{natural} scales provided by
kinematics, thus tracing the path towards possible precision studies of cross
sections.
We limit ourselves to the calculation of azimuthal coefficients up to $n=3$. The extension of our study to higher azimuthal coefficients makes possible to construct other observables, such as angular distributions in $\cos \varphi$, as suggested in Ref.~\cite{Cipriano:2013ooa}.

\begin{figure}[b]
\centering
\includegraphics[scale=0.505,clip]{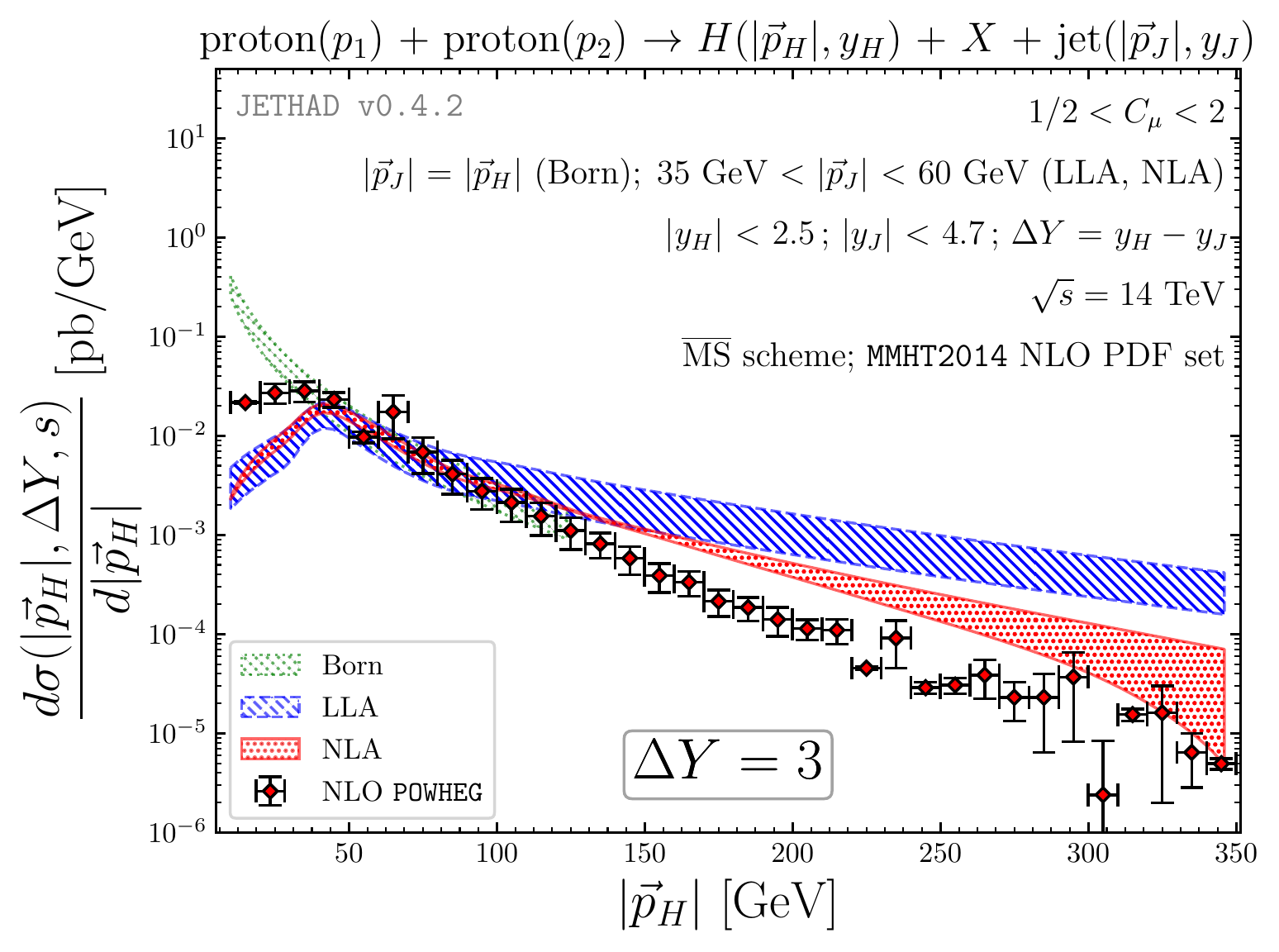}
\includegraphics[scale=0.505,clip]{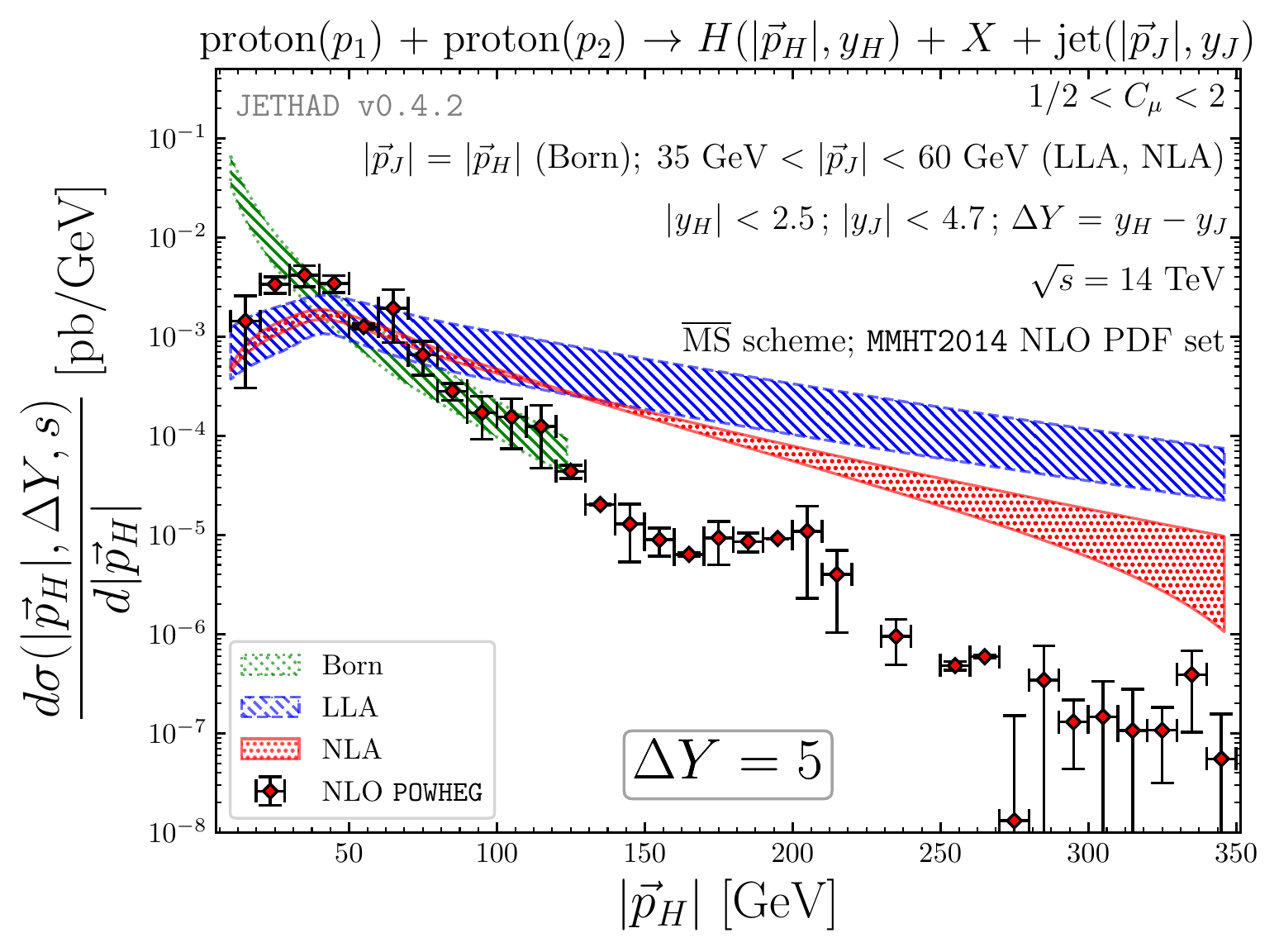}
\caption{$p_T$-dependence of the cross section for the inclusive Higgs-jet
  hadroproduction for 35 GeV $< |\vec p_J| <$ 60 GeV,
  $\sqrt{s} = 14$ TeV and for
  $\Delta Y = 3, 5$. Shaded bands give the combined effect of the scale
  variation with the uncertainty coming from the phase-space numerical
  integration.}
\label{fig:pT_kt-w}
\end{figure}

In Fig.~\ref{fig:pT_kt-w} we present predictions for the $p_H$-distribution,
${d\sigma}/{(d|\vec p_H| d\Delta Y)}$, in the range 10 GeV
$< |\vec p_H| < 2 M_t$, and for two values of the rapidity interval,
$\Delta Y =$ 3, 5.
Here, the Born contribution (green) corresponds to the so-called
\emph{two-gluon} approximation, which describes the back-to-back emission of
the Higgs and of the jet with no additional gluon radiation. From the analytic
point of view, one has
\begin{equation}\nonumber
 \frac{d\sigma^{\rm Born}(|\vec p_H|, \Delta Y, s)}{d|\vec p_H| d\Delta Y} =
 \pi \frac{e^{\Delta Y}}{s} M_{H,\perp}
 \int_{y^{\rm min}_H}^{y^{\rm max}_H}dy_H
 \int_{y^{\rm min}_J}^{y^{\rm max}_J}dy_J
 \, \delta \left( y_H - y_J - \Delta Y \right)
\end{equation}
\begin{equation} \label{pT_distribution_Born}
  \times \alpha_s^2(\mu_{R_1}) \,
\frac{1}{v^2} \frac{|\mathcal{F}(\vec p_H^{\: 2})|^2}
 {128\pi^{3}\sqrt{2(N^2_{c}-1)}} f_{g}(x_H,\mu_{F_1}) 
\end{equation}
\begin{equation}\nonumber
\times  \alpha_s(\mu_{R_2}) \, 2 \sqrt{\frac{C_F}{C_A}} 
\left( \frac{C_A}{C_F} f_{g}(x_J,\mu_{F_2}) + \sum_{a = q,\bar{q}}f_{a}(x_J,\mu_{F_2}) \right)
 \, .
\end{equation}
Our calculation in the Born limit at $\Delta Y = $ 3 (left panel of
Fig.~\ref{fig:pT_kt-w}) is
in fair agreement with the corresponding pattern in Ref.~\cite{DelDuca:2003ba}
(solid line in the left panel of Fig.~2), up to a factor two, due to
the fact that we restricted $\Delta Y$ to be positive, which means that
the Higgs particle is always more forward than the jet\footnote{Note that
  in Ref.~\cite{DelDuca:2003ba} the Higgs
  mass is a free parameter. We compare our result with the corresponding one
  at $M_H =$ 120 GeV.}. In our study, this calculation cannot exceed a given
upper cut-off in the $|\vec p_H|$-range, say around 125 GeV. This is due to our
choice for the final-state kinematic ranges, where consistency with
experimental cuts in the rapidities of the detected objects would lead to
$x_J > 1$ for sufficiently large jet transverse momenta.

Both the LLA (blue) and the NLA (red) series of Fig.~\ref{fig:pT_kt-w} show a
peak (not present in the Born case) at $|\vec p_H|$ around 40 GeV for the two
values of $\Delta Y$, and a decreasing behavior at large $|\vec p_H|$.
For the sake of simplicity, we distinguish three kinematic subregions. The
low-$|\vec p_H|$ region, {\emph{i.e.}} $|\vec p_H| < 10$ GeV, has been excluded
from our analysis, since it is dominated by large transverse-momentum
logarithms, which call for the corresponding all-order
resummation~\cite{TM_resum}, not accounted by our formalism.  To the
intermediate-$|\vec p_H|$ region the set of configurations
where $|\vec p_H|$ is of the same order of $|\vec p_J|$, which ranges from 35
to 60 GeV, corresponds. It is essentially the peak region plus the first part
of the decreasing tail, where NLA bands are totally nested inside the LLA ones.
Here, the impressive stability of the perturbative series unambiguously
confirms the validity of our description at the hand of the BFKL resummation.
Finally, in the large-$|\vec p_H|$ region represented by the long tail,
NLA distributions decouple from LLA ones and exhibit an increasing sensitivity
to scale variation. Here, DGLAP-type logarithms together with
\emph{threshold} effects~\cite{threshold_resum} start to become relevant, thus
spoiling the convergence of the high-energy series.
In Fig.~\ref{fig:pT_kt-w} we present also the
  $p_H$-distributions at $\Delta Y=3$ and$~5$, as obtained by a fixed-order
  NLO calculation through the POWHEG method~\cite{POWHEG}, by suitably
  adapting the subroutines dedicated to the inclusive Higgs plus jet final
  state~\cite{POWHEG_Higgs-jet}. It is interesting to observe that, both
  at $\Delta Y=3$ and $\Delta Y=5$, the NLO fixed-order prediction is
  systematically lower than the LLA- and NLA-BFKL ones and this is more
  evident at the larger $\Delta Y$, where the effect of resummation is
  expected to be more important. This observation provides with an interesting
  window for discrimination between fixed-order and high-energy-resummed
  approaches.

\begin{figure}[b]
\centering
\includegraphics[scale=0.505,clip]{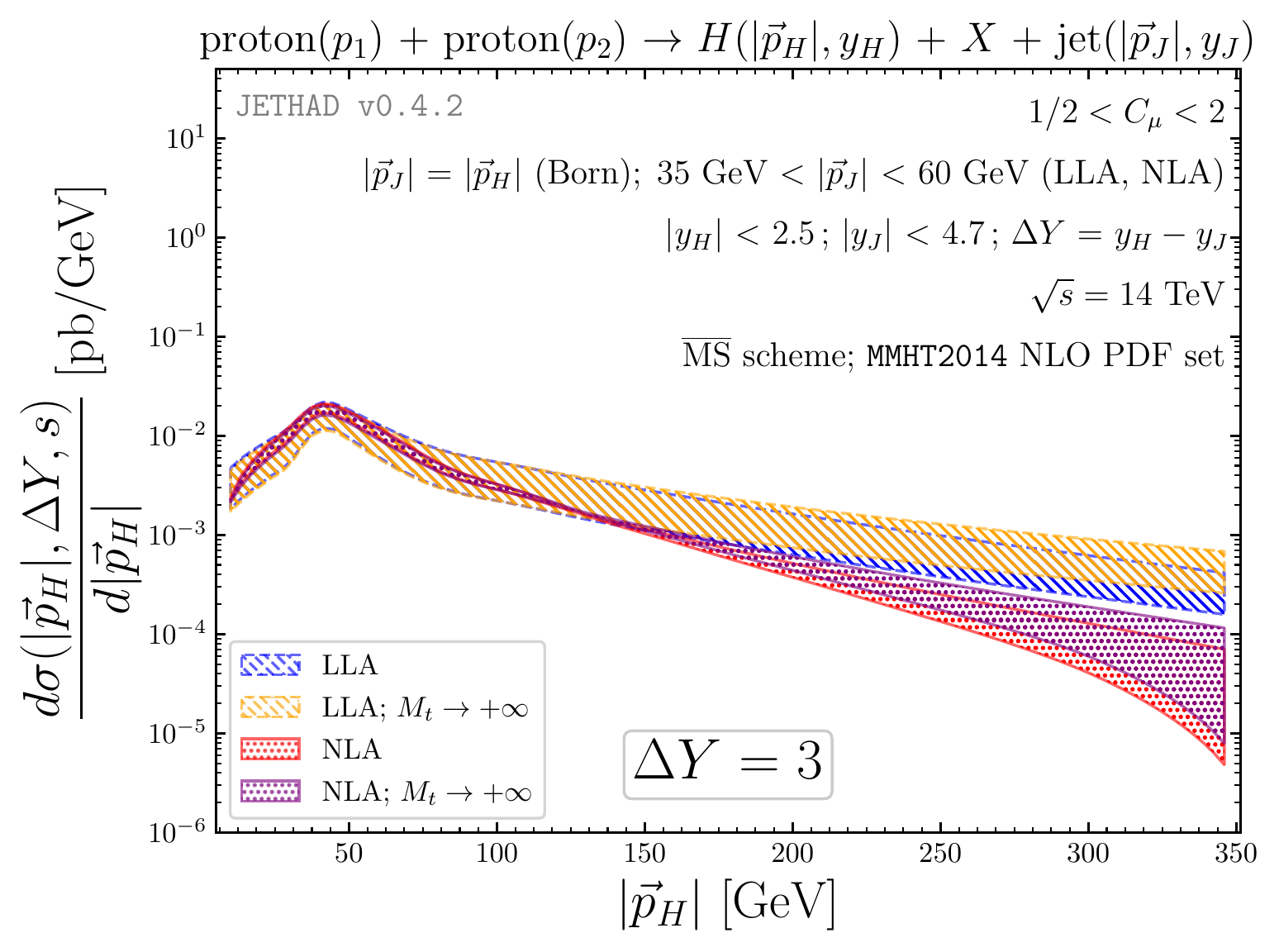}
\includegraphics[scale=0.505,clip]{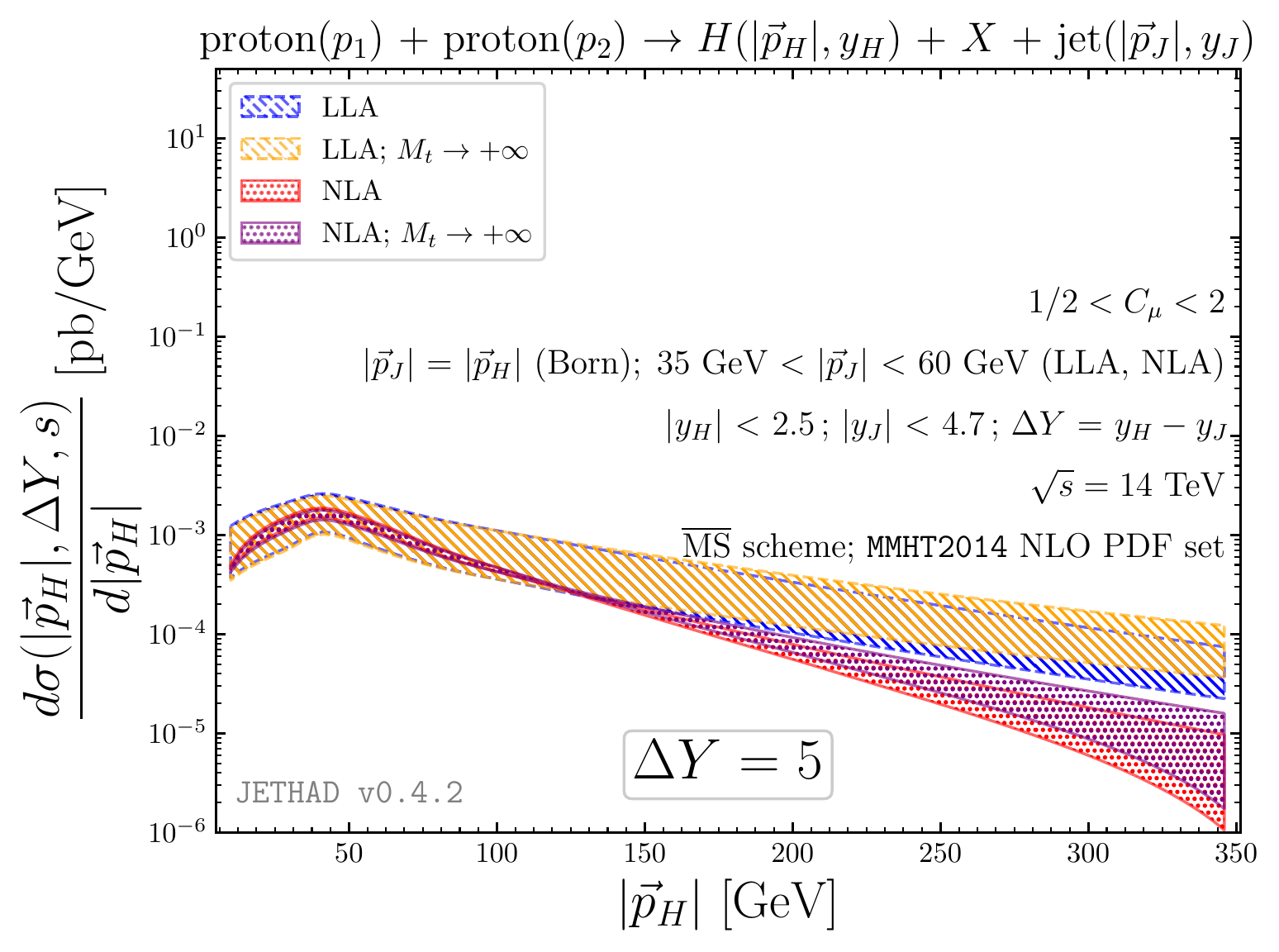}
\caption{$p_T$-dependence of the cross section for the inclusive Higgs-jet
  hadroproduction in the standard case and in the large top-mass limit, 
  for 35 GeV $< |\vec p_J| <$ 60 GeV, $\sqrt{s} = 14$ TeV and for
  $\Delta Y = 3, 5$. Shaded bands give the combined effect of the scale
  variation with the uncertainty coming from the phase-space numerical
  integration.}
\label{fig:pT_kt-w_std-vs-ltop}
\end{figure}

Finally, we compared the distributions presented above with the
  corresponding ones obtained in the large top-mass limit, $M_t \to + \infty$
  (see Eq.~\ref{IF_partonic_mass_limit}). 
  We noted that, when this limit is taken, cross sections
  become at most $5 \div 7 \, \%$ larger,
  whereas the effect on azimuthal correlations
  is very small or negligible. We do not show figures related with this
  comparison, since the bands related to the large top-mass limit are hardly
  distinguishable from the ones with physical top mass.
  The impact on the $p_H$-distribution (Fig.~\ref{fig:pT_kt-w_std-vs-ltop}) is
  also quite small in the $|\vec p_H| \sim |\vec p_J|$ range, while it become
  more manifest when the value of $|\vec p_H|$ increases.

All these considerations brace the message that an exhaustive study of the
$|\vec p_H|$-distribution would rely on a unified formalism where distinct
resummations are concurrently embodied. In particular, the impact of the BFKL
resummation could depend on the delicate interplay among the Higgs transverse
mass, the Higgs transverse momentum and the jet transverse momentum entering,
in logarithmic form, the expressions of partial NLO corrections to impact
factors (see Eqs.~(\ref{cH1}) and~(\ref{cJ1})). Future studies including full
higher-order corrections will allow us to further gauge the stability of our
calculations. 

%
%

\subsection{Numerical specifics and uncertainty estimation}
\label{numerical_strategy}

All the numerical studies were completed making use of \textsc{Jethad}, a
\textsc{Fortran2008-Python3} hybrid library under development at our Group,
which has been recently employed in the analysis of hadron-jet
correlations~\cite{Bolognino:2018oth} and of the inclusive heavy-flavored
jet-pair hadroproduction~\cite{Bolognino:2019yls}. An auxiliary, independent
\textsc{Mathematica} interface allowed us to test the numerical reliability
of our results.
Quark and gluon PDFs were calculated through the \textsc{MMHT2014} NLO PDF
set~\cite{Harland-Lang:2014zoa} as provided by the LHAPDFv6.2.1
interpolator~\cite{Buckley:2014ana}, whereas we selected a two-loop
running coupling setup with $\alpha_s\left(M_Z\right)=0.11707$ and with
dynamic-flavor threshold.

The two relevant sources of numerical uncertainty respectively come from the
multidimensional integration over the final-state phase space (together with
the oscillatory $\nu$-distribution) and from the one-dimensional integral over
the longitudinal momentum fraction $\zeta$ in the NLO impact factor
corrections (Eqs.~(\ref{cH1}) and~(\ref{cJ1})).
They were directly estimated by the \textsc{Jethad} integration tools.
Other potential uncertainties, as the upper cutoff in the numerical
integrations over $|\vec p_H|$, $|\vec p_J|$ and the $\nu$-variable, turned
out to be negligible with respect to the first ones.

Furthermore, we gauged the effect of concurrently varying the renormalization
scales ($\mu_{R_{1,2,c}}$) and the factorization ones ($\mu_{F_{1,2}}$) of them
around their \emph{natural} values in the range 1/2 to two. The parameter
$C_{\mu}$ entering the inset of panels in
Figs.~\ref{fig:C0_kt-asw}, \ref{fig:Rnm_kt-s}, \ref{fig:Rnm_kt-a},
\ref{fig:Rnm_kt-w} and \ref{fig:pT_kt-w} gives the ratio
\begin{equation}
 \label{Cmu}
 C_\mu = 
 \frac{\mu_{{R,F}_1}}{M_{H,\perp}} = 
 \frac{\mu_{{R,F}_2}}{|\vec p_J|} = 
 \frac{\mu_{R_c}}{\sqrt{M_{H,\perp} |\vec p_J|}} \, .
\end{equation}

\section{Conclusions and Outlook}
\label{conclusions}

We have proposed the inclusive hadroproduction of a Higgs boson and of a jet
featuring high transverse momenta and separated by a large rapidity distance as
a new diffractive semi-hard channel to probe the BFKL resummation. 
Statistics for cross sections differential in rapidity, tailored on different
configurations for transverse-momentum ranges at CMS, is encouraging.
At variance with previous analyses, where other kinds of final states were
investigated, cross sections and azimuthal correlations for the Higgs-jet
production exhibit quite a fair stability under higher-order corrections.
Future analyses are needed in order to gauge the feasibility of precision
calculations of the same observables.
We have extended our study to distributions differential in the Higgs
transverse momentum, providing evidence that a high-energy treatment is valid
and can be afforded in the region where Higgs $p_T$ and the jet one are of the
same order.

An obvious extension of this work consists in the full NLA BFKL analysis,
including a NLO jet impact factor, with a realistic implementation of
the jet selection function, and the NLO Higgs impact factor, when
available.

\section*{Acknowledgments}

We thank V.~Bertone, G.~Bozzi and L.~Motyka for helpful discussions
and F.~Piccinini, C.~Del Pio for help on the use of
  the POWHEG code.
FGC acknowledges support from the Italian Ministry of Education, Universities
and Research under the FARE grant ``3DGLUE'' (n. R16XKPHL3N) and from the
INFN/NINPHA project.
MMAM and AP acknowledge support from the INFN/QFT@colliders project.
The work of DI was carried out within the framework of the
state contract of the Sobolev Institute of Mathematics (Project No. 0314-2019-0021)

\end{document}